\documentclass[12pt, a4paper]{article}
\pdfoutput=1

\usepackage[height=25cm,width=16cm]{geometry}
\usepackage{amsfonts}
\usepackage{amsmath}
\usepackage{amssymb}
\usepackage{ascmac}
\usepackage{dcolumn}
\usepackage{bm,here}
\usepackage{comment}
\usepackage{ifpdf}
\usepackage{slashed}
\usepackage{colortbl}
\usepackage{color}
\usepackage[normalem]{ulem}
\usepackage[mathscr]{eucal}
\usepackage{cite}
\usepackage[sort&compress, numbers, merge]{natbib}
\usepackage[utf8]{inputenc}
\usepackage{cancel}
\usepackage{graphicx}
\usepackage{subcaption}
\usepackage{url}
\usepackage{comment}
\usepackage{ifthen}
\usepackage[colorlinks=true, linkcolor=blue, citecolor=blue, urlcolor=black]{hyperref} 

\usepackage{multicol}
\definecolor{red}{rgb}{1,0,0}
\definecolor{ppink}{rgb}{0.921545,0.440586,0.687243}
\definecolor{bblue}{rgb}{0.400000,0.400000,1.000000}
\usepackage[charter]{mathdesign}

\begin{document}

\begin{titlepage}

\begin{flushright}
	IPMU17-0162\\
\end{flushright}

\vskip 3.0cm
	\begin{center}
	{\Large \bf
	Indirect Probe of Electroweakly Interacting Particles \\
	\vskip 0.2cm
	at the High-Luminosity Large Hadron Collider
	}

\vskip 2.0cm
	Shigeki Matsumoto,
	Satoshi Shirai and
	Michihisa Takeuchi

\vskip 0.5cm
	{\it
	Kavli Institute for the Physics and Mathematics of the Universe (WPI), \\
	The University of Tokyo Institutes for Advanced Study, \\
	The University of Tokyo, Kashiwa 277-8583, Japan
	}

\vskip 5.0cm
	\abstract{Many extensions of the standard model (SM) involve new massive particles charged under the electroweak gauge symmetry. The electroweakly interacting new particles affect various SM processes through radiative corrections. We discuss the possibility of detecting such new particles based on the precise measurement of the SM processes at high energy hadron colliders. It then turns out that Drell-Yan processes receive radiative corrections from the electroweakly interacting particles at the level of ${\cal O}$(0.1--10)\%. It is hence possible to indirectly search for the Higgsino up to the mass of 400\,GeV and the quintet (5-plet) Majorana fermion up to the mass of 1200\,GeV at the high-luminosity running of the Large Hadron Collider, if the systematic uncertainty associated with the estimation of the SM background becomes lower than the statistical one.}

\end{center}
\end{titlepage}

\section{Introduction}
\label{sec: intro}
Various types of electroweakly interacting massive particles (EWIMPs) are introduced in new physics models beyond the standard model (SM). For instance, in supersymmetric (SUSY) extensions of the SM, almost all supersymmetric partners of SM particles are charged under the electroweak symmetry, namely SU(2)$_{\rm L}$ $\times$ U(1)$_{\rm Y}$ gauge symmetry. This is also the case for other new physics models such as extra dimension models, extended Higgs models and so on. Such EWIMPs often play significant roles for the origin of the electroweak symmetry breaking and/or can be excellent candidates for dark matter in our universe.

The latter role is particularly interesting, as the electroweak interaction makes the dark matter satisfying the weakly interacting massive particle (WIMP) hypothesis and visible at direct and indirect dark matter detection experiments. One of such candidates is the wino dark matter in SUSY having the electroweak quantum number of ${\bf 3}_0$, and it has actually rich dark matter signatures thanks to the interaction\,\cite{Hisano:2003ec, *Hisano:2004ds, *Hisano:2005ec, Hisano:2010fy, *Hisano:2010ct, *Hisano:2012wm, *Hisano:2015rsa}. It is also worth pointing out that the wino dark matter is motivated very well from the viewpoint of particle physics theory; it is a generic prediction from SUSY SMs with the anomaly mediation\,\cite{Randall:1998uk, Giudice:1998xp}. This framework is known to be compatible with the so-called ``mini-split SUSY'' scenario\,\cite{Wells:2003tf, *Wells:2004di, ArkaniHamed:2004fb, *Giudice:2004tc, *ArkaniHamed:2004yi, *ArkaniHamed:2005yv}, and the discovery of the 125\,GeV Higgs boson\,\cite{Aad:2012tfa, Chatrchyan:2012ufa} triggers the framework to attract more and more attention\,\cite{Hall:2011jd, *Hall:2012zp, *Nomura:2014asa, Ibe:2011aa, *Ibe:2012hu, Arvanitaki:2012ps, ArkaniHamed:2012gw}. In fact, detailed phenomenological studies on the wino dark matter are stimulated in many studies\,\cite{Cohen:2013ama, Fan:2013faa, Bhattacherjee:2014dya, Ibe:2015tma}. Another interesting candidate is the dark matter having a large electroweak quantum number, because such a quantum number makes the dark matter particle stable without imposing any ad hoc dark matter parity. The quintet fermion whose quantum number is ${\bf 5}_0$ and the septet scalar having the quantum number of ${\bf 7}_0$ are such examples. Those are now referred as the ``minimal dark matter''\,\cite{Cirelli:2005uq, *Cirelli:2007xd, *Cirelli:2009uv}. There also be an interesting candidate motivated from the electroweak symmetry breaking; the Higgsino having the electroweak quantum number of ${\bf 2}_{\pm 1/2}$, which is often predicted to be the lightest SUSY particle in the so-called natural SUSY scenario\,\cite{Kitano:2005wc}. Interestingly, its mass is required to be smaller than 350\,GeV to obtain the electroweak scale naturally\,\cite{Baer:2012up, Baer:2013gva}, which is kinematically accessible at the Large Hadron Collider (LHC). Discovery and measurement of EWIMP at colliders are thus a crucial test for physics beyond the SM.

Collider signals of EWIMP is strongly model-dependent. For instance, it is crucial whether or not the EWIMP is also charged under the SU(3)$_{\rm C}$ symmetry for its production at hadron colliders. Moreover, collider signatures are strongly affected by whether and how the EWIMP decays. The search for the EWIMP at hadron colliders becomes difficult in general if its decay products are very soft\,\cite{Gori:2013ala, Han:2013usa, Han:2014kaa, Bramante:2014dza, Baer:2014kya, Bramante:2014tba, Ismail:2016zby}. A prominent example is the direct production of the electroweakly interacting dark matter. Though the search for the large missing energy accompanied with high $P_{\rm T}$ jets or photons is conventionally used to search for the EWIMP, it does not work efficiently due to huge SM backgrounds as well as the small production cross section of the EWIMP. In some cases, the mass difference among the SU(2)$_{\rm L}$ multiplet becomes so small that its charged component is long-lived. When the charged track caused by the long-lived particle is detectable, it may overcome the huge SM backgrounds.\footnote{In fact, the most sensitive search for the pure wino dark matter is based on this strategy\,\cite{Ibe:2006de, *Buckley:2009kv, Asai:2007sw, *Asai:2008sk, *Asai:2008im, ATLAS:2017bna}, as the charged wino has a decay length of about 6\,cm\,\cite{Ibe:2012sx, McKay:2017rjs}. Furthermore, it has been shown recently that a much shorter decay length becomes detectable by improving the track reconstruction and/or modifying the detector\,\cite{Mahbubani:2017gjh, Fukuda:2017jmk}. It is then even possible to detect the almost pure Higgsino whose charged component has a decay length shorter than 1\,cm. The strategy based on the charged track is also applicable to the case that EWIMP is not dark matter but a coannihilating partner like the gaugino coannihilation\,\cite{ArkaniHamed:2006mb, Harigaya:2014dwa, Bharucha:2017ltz}; the search for the EWIMP becomes powerful despite its soft decay products when it is long-lived enough\,\cite{Nagata:2015pra}} This is, however, not a generic feature of EWIMP. Even if we consider EWIMP dark matter, its coupling to the Higgs field may enhance the mass difference, making the decay length of the charged component too short\,\cite{Hisano:2014kua, Nagata:2014wma, Nagata:2014aoa}. It is therefore important to develop an independent method for the EWIMP search which does not rely on the charged track search.

We consider an indirect probe of EWIMP through its radiative corrections to SM processes. We particularly focus on the dilepton production by the Drell-Yan process at the LHC; the EWIMP is expected to modify the lepton invariant mass distribution ($m_{\ell \ell}$). It is known that, when $m_{\ell \ell}$ is much smaller than twice the EWIMP mass ($m$), the correction is effectively described by dimension-six operators and is proportional to $m_{\ell \ell}^2/m^2$. On the other hand, the EWIMP affects the running of the electroweak gauge couplings when $m_{\ell \ell} \gg 2 m$, leading to the correction proportional to $\ln (m_{\ell \ell}^2/m^2)$\,\cite{Alves:2014cda, Gross:2016ioi, Farina:2016rws}. In this paper, we show that the correction becomes an extremum of ${\cal O}$(0.1--10)\% when $m_{\ell \ell} \simeq 2m$, making the EWIMP detectable at the LHC utilizing this feature. We find that the present observation of the dilepton mass distribution already excludes the EWIMP having a large electroweak quantum number. It will be possible to search for the EWIMP having a smaller quantum number at the high-luminosity LHC (HL-LHC), so that the indirect EWIMP search becomes as important as the direct production search with mono-$X$ and missing energy in the future.

\section{Radiative correction from EWIMP}
\begin{figure}[t]
	\centering
	\includegraphics[clip, width = 0.5 \textwidth]{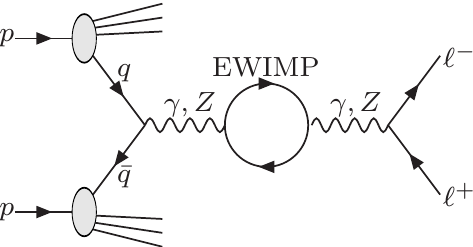} 
	\caption{\it \small Radiative correction from the electroweakly interacting massive particle (EWIMP).}
	\label{fig: FG}
\end{figure}

We focus on the Drell-Yan process $p p \to \ell^+ \ell^- + X$ in this paper with $p$ and $\ell^\pm$ being a proton and a lepton, and discuss how EWIMP modifies the lepton invariant mass distribution $m_{\ell \ell}$ at high energy hadron colliders. The EWIMP affects the differential cross section of the process through the loop correction shown in Fig.\,\ref{fig: FG}. After integrating the EWIMP field out from the original Lagrangian at one-loop level, we obtain the following effective Lagrangian:\footnote{We only take the electroweak gauge interactions of the EWIMP into account for simplicity and neglect other renormalizable interactions (such as the Yukawa interaction, etc.) to derive the effective Lagrangian.}
\begin{eqnarray}
	\mathcal{L}_\mathrm{eff} = \mathcal{L}_\mathrm{SM}
	+ \kappa \frac{g^2 C_{WW}}{8} W^a_{\mu\nu}\,\Pi(-D^2/m^2)\,W^{a \mu \nu}
	+ \kappa \frac{g^{\prime 2} C_{BB}}{8} B_{\mu\nu}\,\Pi(-\partial^2/m^2)\,B^{\mu \nu}
	+ \cdots\,,
	\label{eq: effective lagrangian}
\end{eqnarray}
where $\mathcal{L}_\mathrm{SM}$ is the SM Lagrangian, $m$ is the EWIMP mass, $g$ ($g'$) is the gauge coupling of SU(2)$_{\rm L}$ (U(1)$_{\rm Y}$) and $W^a_{\mu\nu}$ ($B_{\mu\nu}$) is the corresponding field strength tensor, respectively, with $D$ being the covariant derivative acting on $W^{a \mu \nu}$. Parameters $C_{WW}$ and $C_{BB}$ are defined as \mbox{\boldmath $C_{WW} \equiv \xi (n^3 - n)/6$} and \mbox{\boldmath $C_{BB} \equiv 2 \xi n Y^2$} for the SU(2)$_{\rm L}$ n-tuplet EWIMP having the hypercharge $Y$ and the color degree of freedom $\xi$, while $\kappa$ takes a value of 1/2, 1, 4 and 8 when the EWIMP is a real scalar, complex scalar, Majorana and Dirac fermions, respectively. The ellipsis in the Lagrangian includes operators composed of the strength tensors more than two, but those are not relevant to the following discussion. The function $\Pi(x)$ is the renormalized self-energy of the electroweak gauge bosons from the EWIMP loop:
\begin{eqnarray}
	\Pi(x) = 
	\begin{cases}
	\dfrac{1}{16\pi^2} \displaystyle\int_0^1 dy\,y (1 - y) \ln [1 - y (1 - y)\,x - i 0^+] & (\mbox{Fermion}), \\
	\dfrac{1}{16\pi^2} \displaystyle\int_0^1 dy\,(1 - 2y)^2 \ln [1 - y (1 - y)\,x - i 0^+] & (\mbox{Scalar}).
	\end{cases}
\end{eqnarray}
We have used the $\overline{\mathrm{MS}}$ regularization scheme with the renormalization scale of $\mu = m$.

All the effect of the EWIMP on the process is encoded in the operators involving two field strength tensors. When the EWIMP mass is much larger than the partonic collision energy $\hat{s}^{1/2}$, namely $m^2 \gg -\partial^2 = \hat{s}$, the operators give dimension-six ones, $(D_\mu W^{a \mu \nu})(D^\rho W^a_{\rho \nu})$ and $(\partial_\mu B^{\mu \nu})(\partial^\rho B_{\rho \nu})$. In contrast, the function $\Pi$ behaves as $\Re [ \Pi(-\partial^2/m^2) ] \sim \log(\hat{s}/m^2)$ when $m^2 \ll -\partial^2 = \hat{s}$, which is eventually translated into the running effect of the electroweak gauge couplings. The effect of the EWIMP in these two extreme regions has already been studied in several papers\,\cite{Alves:2014cda, Gross:2016ioi, Farina:2016rws}. On the other hand, we use the effective Lagrangian in eq.\,(\ref{eq: effective lagrangian}) directly, as we are interested in the effect at the region around $s \sim 4 m^2$.

The matrix element of the Drell-Yan process is obtained from the effective Lagrangian. Leading order (LO) contribution is from SM interactions and its explicit form is
\begin{eqnarray}
	{\cal M}_{\rm LO} [ q(p) \bar{q}(p') \to \ell^-(k) \ell^+(k') ] =
	\sum_{V = \gamma,\,Z}
	\frac{[\bar{v}(p') \gamma^\mu \Gamma^V_q u(p)] [\bar{u}(k) \gamma_\mu \Gamma^V_l v(k')] }
	{\hat{s} - m^2_V },
	\label{eq: LO}
\end{eqnarray}
where $\Gamma^Z_f = g_Z (v_f - a_f \gamma_5)$ and $\Gamma^\gamma_f = e Q_f$, while $g_Z = g/c_W$ and $e = g s_W$, where $s_W = \sin \theta_W$ and $c_W = \cos \theta_W$ with $\theta_W$ being the weak mixing angle. Coefficients $(v_f,\,a_f,\,Q_f)$ are $(1/4 - 2 s_W^2/3,\,1/4,\,2/3)$, $(-1/4 + s_W^2/3,\,-1/4,\,-1/3)$ and $(-1/4 + s_W^2,\,-1/4,\,-1)$ for up-type quarks, down-type quarks and charged leptons, respectively. The mass of the electroweak gauge boson is given by $m_V$, while the center-of-mass energy at this parton-level process is denoted by $\hat{s}^{1/2}$. Next leading order contribution (NLO) to the matrix element from the EWIMP loop diagram shown in Fig.\,\ref{fig: FG} is given by the following formula:
\begin{eqnarray}
	{\cal M}_{\rm BSM} [ q(p) \bar{q}(p') \to \ell^-(k) \ell^+(k') ] = 
	\sum_{V, V'}
	\frac{ d_{VV'} [\bar{v}(p') \gamma^\mu \Gamma^V_q u(p)]
	\,\hat{s}\,\Pi(\hat{s}/m^2)\,
	[\bar{u}(k) \gamma_\mu \Gamma^{V'}_l v(k')] }
	{ (\hat{s} - m^2_V) (\hat{s} - m^2_{V'}) },
	\label{eq: NLO}
\end{eqnarray}
where each coefficient $d_{VV'}$ in the numerator is defined as $d_{ZZ} = \kappa\,(g_Z^2/2)(c_W^4 C_{WW} + s_W^4 C_{BB})$, $d_{\gamma\gamma} = \kappa\,(e^2/2)(C_{WW} + C_{BB})$ and $d_{Z\gamma} = d_{\gamma Z} = \kappa\,(e\,g_Z/2) (c_W^2 C_{WW} - s_W^2 C_{BB})$, respectively.

\begin{figure}[t]
	\centering
	\includegraphics[clip, width = 0.6 \textwidth]{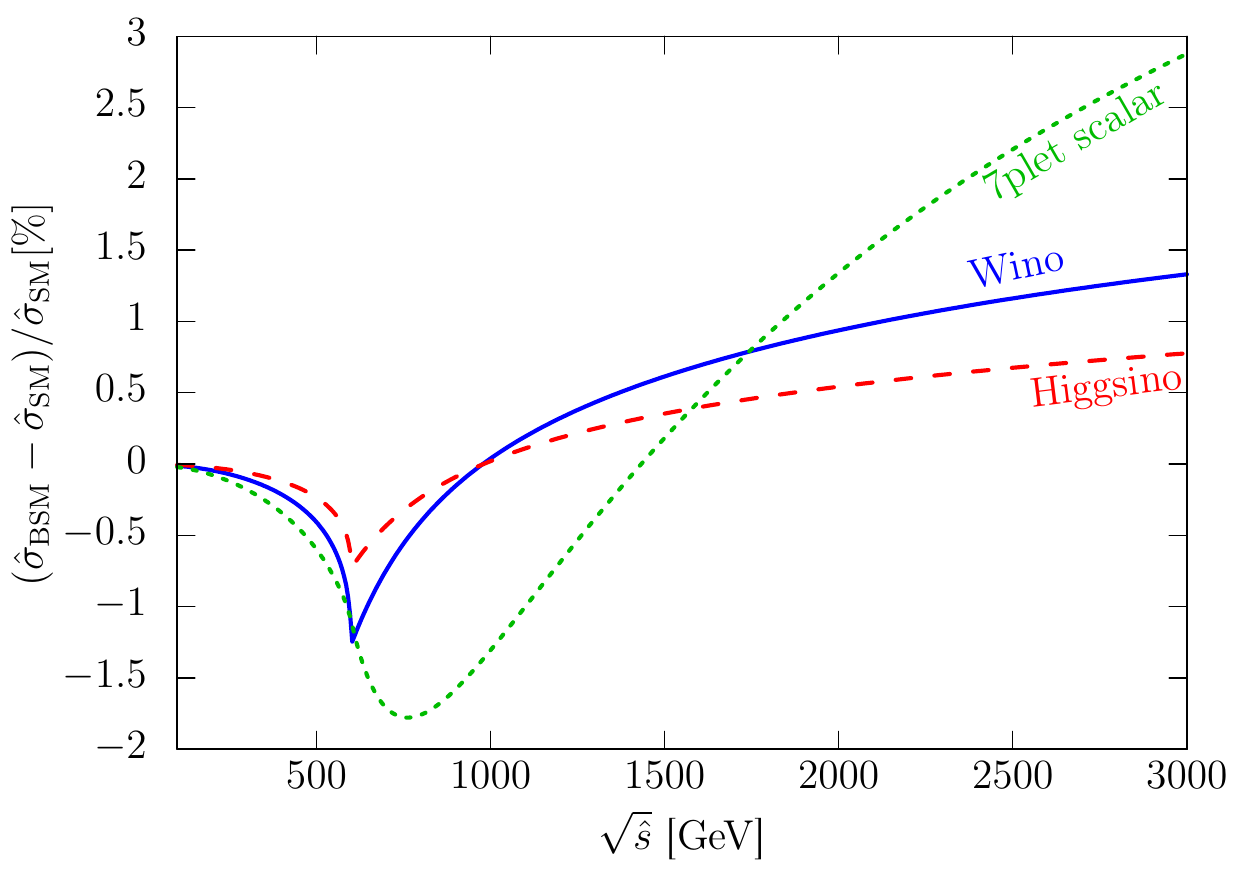} 
	\caption{\sl \small The difference between the differential cross sections of the Drell-Yan process with and without the EWIMP contribution (normalized by the SM prediction) at the 13\,TeV LHC as a function of $m_{\ell \ell} = \hat{s}^{1/2}$. Three EWIMP cases are depicted: Wino (Majorana fermion with the quantum number of $3_0$), Higgsino (Dirac fermion with the quantum number of $2_{\pm 1/2}$) and bosonic minimal dark matter (real scalar with the quantum number of $7_0$) with their masses fixed to be 300\,GeV.}
	\label{fig: example}
\end{figure}

We show in Fig.\,\ref{fig: example} that how the EWIMP contribution modifies the lepton invariant mass distribution at the 13\,TeV LHC, where the difference between the differential cross sections of the Drell-Yan process with and without the EWIMP contribution (normalized by the SM prediction at LO) is shown as a function of the lepton invariant mass $m_{\ell \ell} = \hat{s}^{1/2}$ for three EWIMP cases; Wino (Majorana fermion with the quantum number of $3_0$), Higgsino (Dirac fermion with the quantum number of $2_{\pm 1/2}$) and bosonic minimal dark matter (real scalar with the quantum number of $7_0$) with their masses fixed to be 300\,GeV. The SM cross section $\hat{\sigma}_{\rm SM}$ is calculated at leading order. The cross section of the fermionic minimal dark matter (Majorana fermion with the quantum number of $5_0$)\,\cite{Cirelli:2005uq, Cirelli:2007xd, Cirelli:2009uv} is about five times larger than that of the wino. It can be seen that the modification becomes small when $m_{\ell \ell} \ll m$, while receives a logarithmic correction when $m_{\ell \ell} \gg m$, as expected from the discussion above. On the other hand, when $m_{\ell \ell} \sim 2m$, the correction shows a characteristic feature due to the interference between LO and NLO contributions. The deviation becomes an (almost) extremum at $\hat{s}^{1/2} = 2 m$ for a fermionic (scalar) EWIMP, which is analytically given by
\begin{eqnarray}
	\frac{ \hat{\sigma}_{\rm BSM} - \hat{\sigma}_{\rm SM} } { \hat{\sigma}_{\rm SM}}
	\simeq \Re \left[ \frac{ 2 {\cal M}_{\rm NLO} } { {\cal M}_{\rm LO} } \right]
	\simeq \frac{ \kappa\,g^2\,(n^3 - n)} {6}\,\Pi(4),
\end{eqnarray}
where the effect of gauge boson mass and U(1)$_{\rm Y}$ interaction is neglected. The loop function $\Pi(4)$ takes a value of $-1/(36 \pi^2)$ for a fermionic EWIMP and $-1/(72 \pi^2)$ for a scalar EWIMP. This characteristic feature is expected to be utilized to detect the EWIMP efficiently.

\section{Analysis of Collider Signal}
We discuss here the detection capability of EWIMP by measuring the dilepton distribution at the LHC, where 36\,fb$^{-1}$ data at the 13\,TeV running is mainly used in our analysis\,\cite{Aaboud:2017buh}. We consider two different methods to deal with the ${\cal O}$(0.1--1)\,\% deviation from the SM background. One is the ``fitting based search'', for the EWIMP contributes to the distribution destructively, as we have seen in the previous section. The other one is based on the background estimation through the Monte-Carlo simulation. The simulation now reproduces the observed data very well, so that the EWIMP contribution will be efficiently searched for through the likelihood test of the ``EWIMP signal + SM background'' hypothesis.

\subsection{Fitting based search}
\label{subsec: method 1}

\begin{figure}[t]
	\centering
	\subcaptionbox{\label{fig: ee}Dielectron channel}{\includegraphics[width=0.47\textwidth]{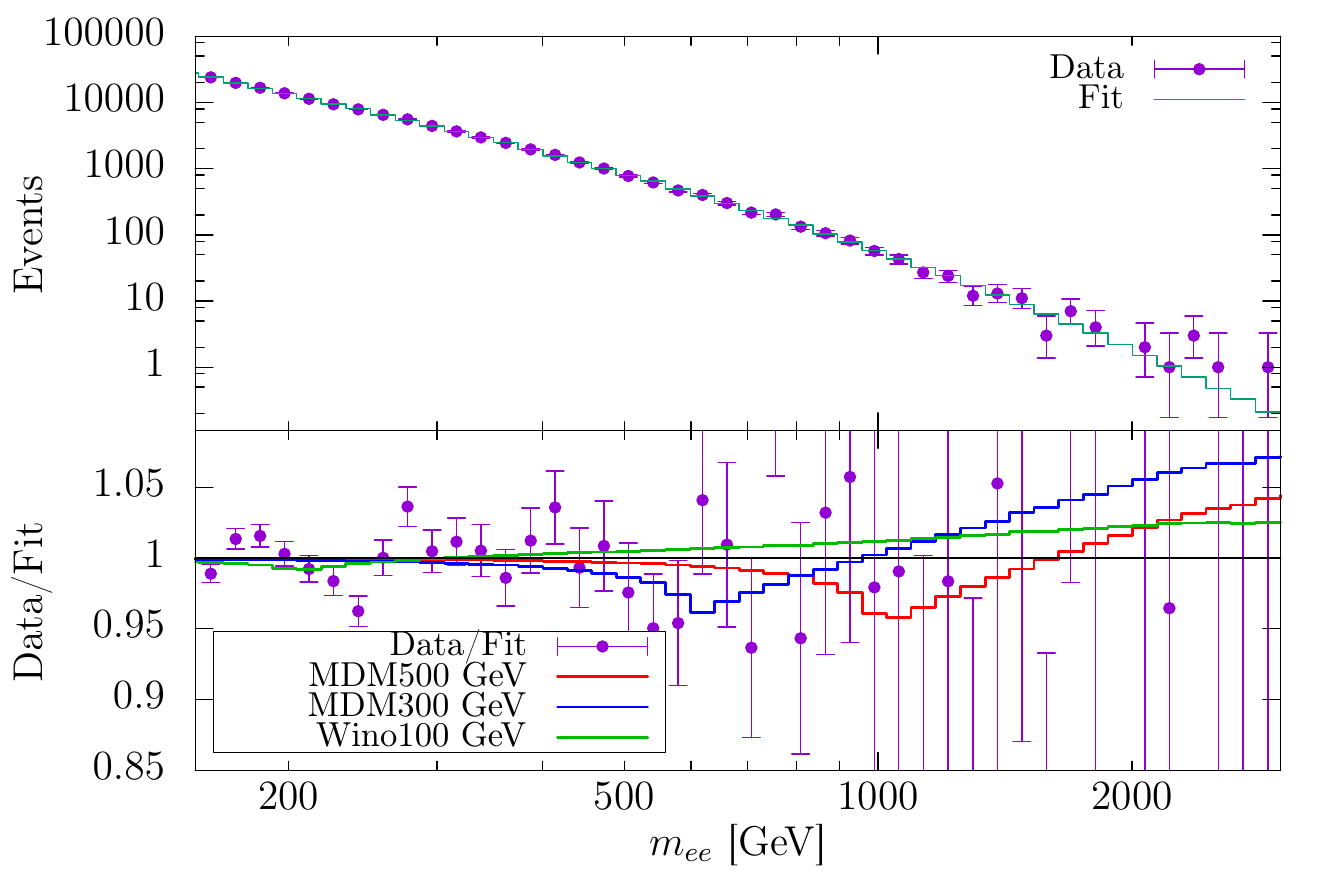}}
	\qquad
	\subcaptionbox{\label{fig: mm}Dimuon channel}{\includegraphics[width=0.47\textwidth]{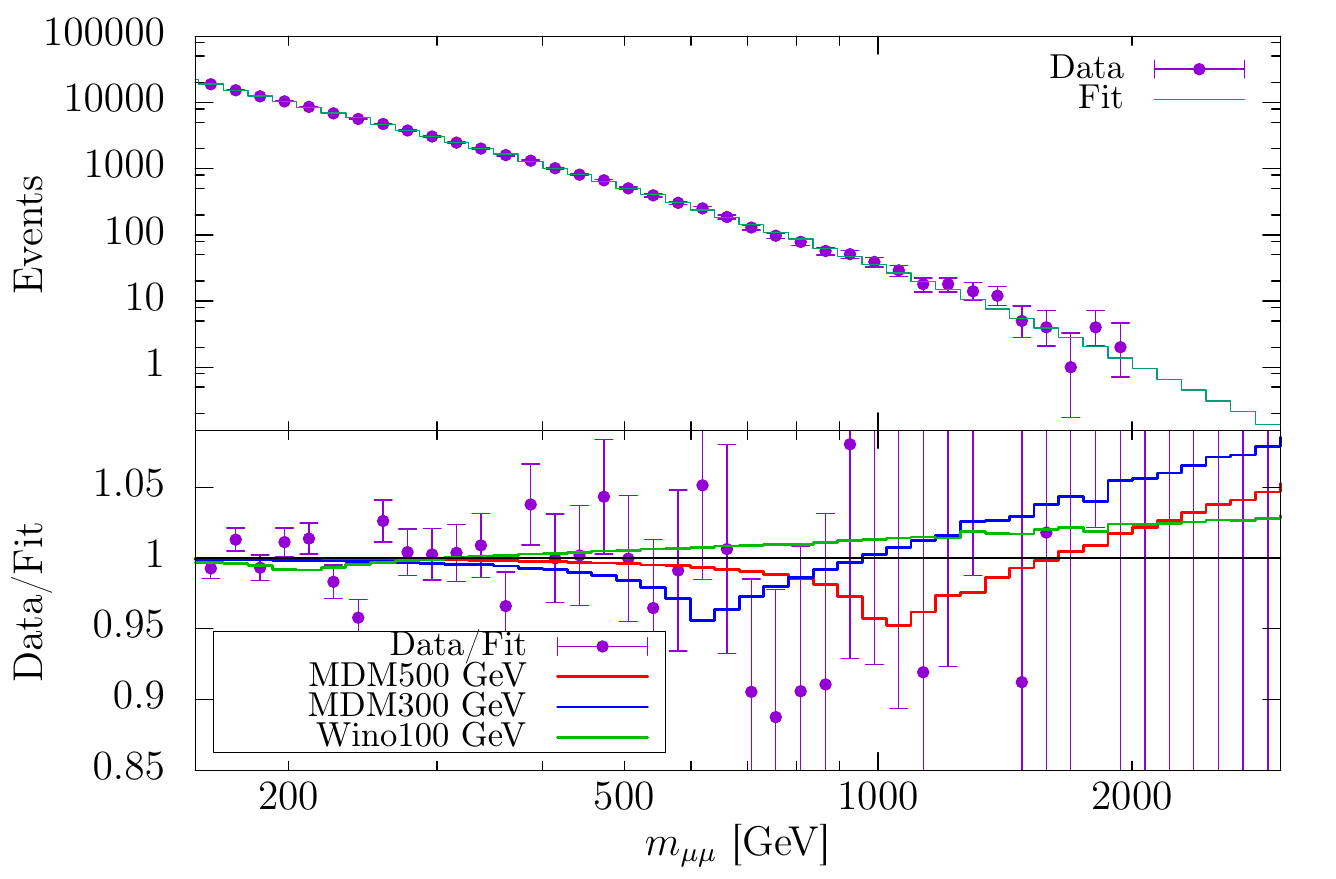}}
\caption{\sl \small Fitting the observational data by the SM background based on the function in eq.\,(\ref{eq: bg}) for (a) dielectron and (b) dimuon channels with $m_{\ell \ell }$ being the lepton invariant mass. The data of 36\,fb$^{-1}$ at the 13\,TeV LHC is used. Bottom panel in each figure shows the ratio between the data and the background as well as expected signals of the wino with the mass of 100\,GeV and the fermionic (5-tuplet) minimal dark matter (MDM) with its mass fixed to be 300\,GeV and 500\,GeV.}
\label{fig: fit}
\end{figure}

The analysis is essentially the same as the conventional bump search at dilepton channels. Since the SM background is expected to give a smooth distribution on the channels and the observed data shows such a smooth distribution too, we can estimate the SM background in a data-driven way, namely by fitting the data using the following function\,\cite{Aaltonen:2008dn}:
\begin{eqnarray}
	\frac{ d N_{\rm BG} } { dm_{\ell \ell} } = p_1\, (1 - x)^{p_2}\,x^{p_3 + p_4\,\log(x) + p_5\,\log^2(x)},
	\label{eq: bg}
\end{eqnarray}
where $N_{\rm BG}$ is the leptonic invariant mass ($m_{\ell \ell}$) distribution of the SM background process, and $x = m_{\ell \ell}/s^{1/2}$ with $s^{1/2} = 13$\,TeV being the center-of-mass energy of the proton collision. The fitting is performed in the region of 150\,GeV $< m_{\ell \ell} <$ 3000\,GeV, and the result is shown in Fig.\,\ref{fig: ee} and Fig.\,\ref{fig: mm} for the dielectron and the dimuon channels, respectively. The background function used in eq.\,(\ref{eq: bg}) is seen to fit the observed data very well.

We then perform the likelihood test of the ``EWIMP signal + SM background'' hypothesis by comparing the observational data shown in Fig.\,\ref{fig: fit} with the following function:
\begin{align}
	\frac{ dN } { dm_{\ell \ell} } = \frac{ dN_{\rm BG} } { dm_{\ell \ell} } + \frac{ dN_{\rm EWIMP} } { dm_{\ell \ell} }.
\label{eq:signal}
\end{align}
The likelihood is calculated in each bin of the lepton invariant mass based on the Poisson distribution, and maximize the total likelihood in the region of 150\,GeV $< m_{\ell \ell} <$ 3000\,GeV. The main contribution to the signal part, $dN_{\rm EWIMP}/dm_{\ell \ell}$, comes from the interference between the LO (SM diagram) and the NLO (EWIMP one-loop diagram). In order to take into account the effect of k-factor, kinematical selection and so on, we calculate $dN_{\rm EWIMP}/dm_{\ell \ell}$ by multiplying the factor $(\hat{\sigma}_{\rm BSM}- \hat{\sigma}_{\rm SM})/\hat{\sigma}_{\rm SM}$ obtained in the previous section to the so-called $Z/\gamma^*$ background number estimated by the ATLAS collaboration\,\cite{Aaboud:2017buh} at each bin.

\subsection{MC based search}
\label{subsec: method 2}

The analysis is almost the same as the previous one, but the SM background is estimated by the Monte-Carlo (MC) simulation, as the ATLAS collaboration adopts. The current systematic uncertainty is a few percent for $m_{\ell \ell} \lesssim 1$\,TeV. Using the MC based background and its systematic uncertainty given by the collaboration, we construct the likelihood for the ``EWIMP signal + SM background'' hypothesis, and put a constraint on the EWIMP. We assume that the systematic uncertainty at each bin is independent of others for simplicity.

\subsection{Capability of EWIMP detection at LHC}

\begin{figure}[t]
	\centering
	\subcaptionbox{\label{fig: fit_result} Fitting based search}{\includegraphics[width=0.47\textwidth]{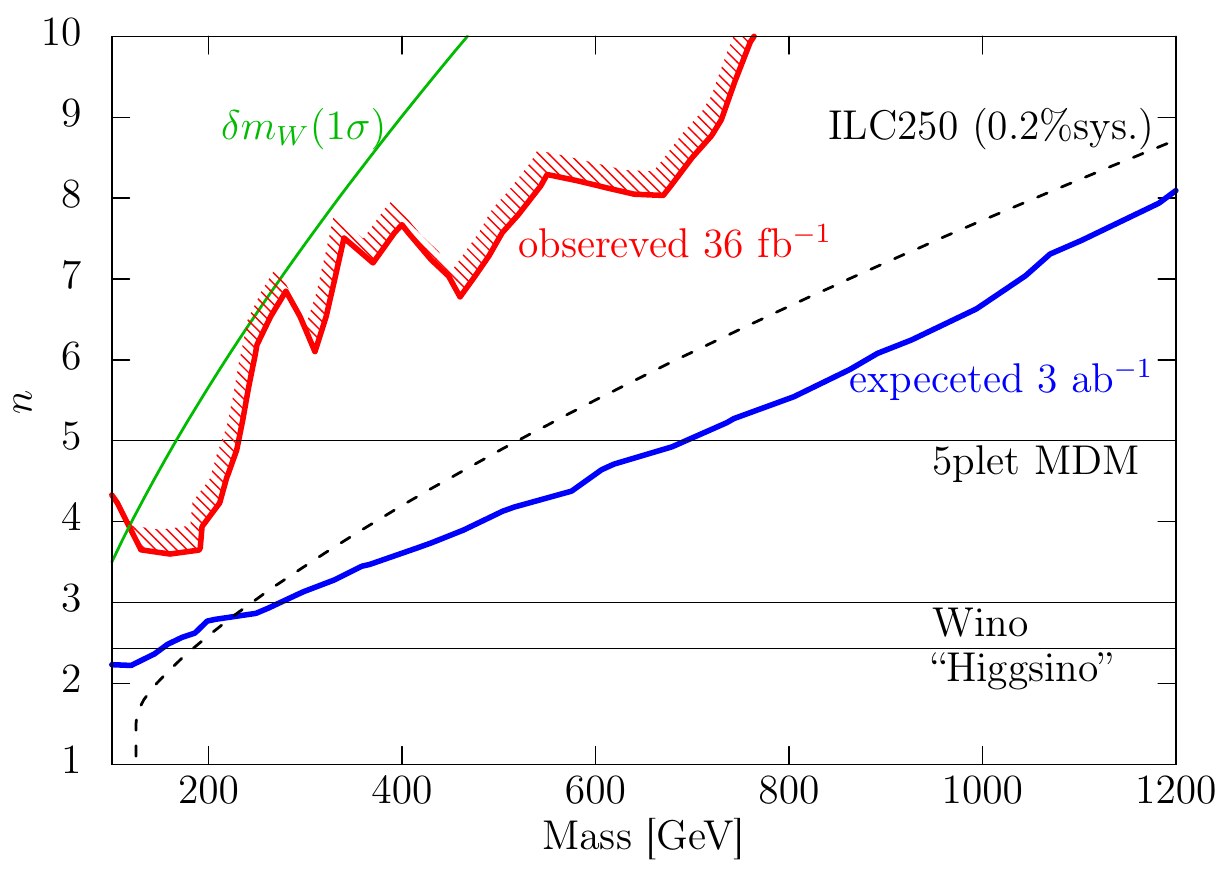}}
	\qquad
	\subcaptionbox{\label{fig: mc_result} MC based search}{\includegraphics[width=0.47\textwidth]{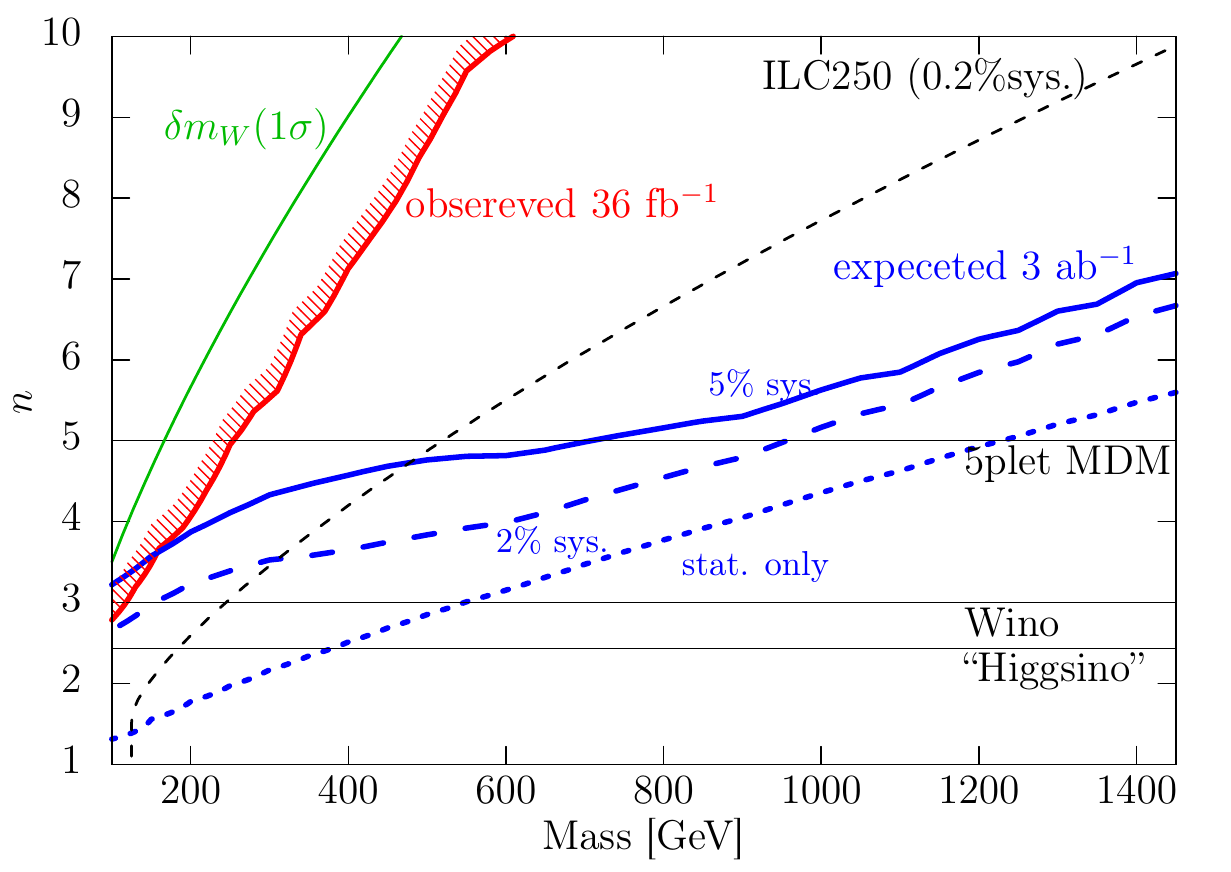}}
	\caption{\sl \small The constraint on a Majorana fermionic EWIMP at 95\% confidence level obtained by the fitting based analysis (left panel) and the MC based analysis (right panel) on the plane of the EWIMP mass and the SU(2)$_{\rm L}$ quantum number `n'. The present constraint from 36\,fb$^{-1}$ data at the 13\,TeV running is shown as a thick red line (associated with a small hatch) in both panels, while the future expected constraint from 3\,ab$^{-1}$ data at the 14\,TeV running is shown as a blue solid line in the left panel and blue solid, dashed and dotted lines in the right panel, depending on the systematic uncertainty associated with the SM background estimation. Regions above the lines are (expected to be) excluded. Theoretical predictions from the well-motivated EWIMP candidates; wino, fermionic minimal dark matter (5plet MDM) and Higgsino, are also shown as horizontal (thin solid) black lines. The green solid line is the constraint from the electroweak precision measurement ($W$ boson mass) at 1$\sigma$ level with the experimental error being 15\,MeV. Future expected constraint from the international linear collider (ILC) at the 250\,GeV running is shown as a black dotted line.}
\label{fig: result}
\end{figure}

We are now at the position to discuss the capability of the EWIMP detection at the LHC. Fig.\,\ref{fig: result} shows the constraint on a Majorana fermionic EWIMP at 95\% confidence level. The SU(2)$_{\rm L}$ quantum number `$n$' is treated as a real number to depict the figure with the hypercharge of the EWIMP being zero. Three horizontal (thin solid) black lines are predictions from the well-motivated EWIMP candidates; wino, fermionic minimal dark matter (5plet MDM) and Higgsino. The SU(2)$_{\rm L}$ quantum number of the Higgsino is estimated to be $n \simeq 2.43$, as it is not a Majorana fermion but a Dirac one. Since the effect of non-zero hypercharge on the Higgsino prediction is negligibly small, it is set to be zero. The constraint obtained by the ``fitting based search'' in Sec.\,\ref{subsec: method 1} is shown in the left panel (Fig.\,\ref{fig: fit_result}), while that obtained by the ``MC based search'' in Sec.\,\ref{subsec: method 2} is shown in the right panel (Fig.\,\ref{fig: mc_result}), respectively.

The present constraint using 36\,fb$^{-1}$ data at the 13\,TeV running is given by a thick red line (associated with a small hatch) in both panels. It can be seen that the fermionic minimal dark matter with the mass below about 250\,GeV is already excluded in both searches. We also consider the future prospect of the EWIMP detection at the HL-LHC assuming 3\,ab$^{-1}$ data at the 14\,TeV running,\footnote{In the fitting based search, the mock data is generated based on the fitting function\,(\ref{eq: bg}), which is corrected by multiplying the ratio between $\hat{\sigma}_{\rm SM}$ at $\sqrt{s} =$ 14\,TeV and 13\,TeV. In the MC based analysis, the mock data is generated based on the background distribution estimated by ATLAS\,\cite{ Aaboud:2017buh} instead of the fitting function.} which is given by a blue solid line in the left panel and blue solid, dashed and dotted lines in the right panel, as the future expected constraint at 95\% confidence level. The systematic uncertainty in the MC based search is set to be 5\% (blue solid line), 2\% (blue dashed line) and 0\% (blue dotted line), respectively.

In the fitting based analysis, it is possible to test Higgsino, wino and fermionic minimal dark matter with their masses up to 150\,GeV, 300\,GeV and 700\,GeV, respectively. On the other hand, in the MC based analysis, it is in principle to test these well-motivated EWIMPs with their masses up to 380\,GeV, 550\,GeV and 1200\,GeV, respectively. This result, of course, depends strongly on how well the systematic uncertainty (associated with the estimation of the SM background) is controlled. The use of the characteristic feature on the dilepton channels is actually very powerful. To see this, we show the constraint on the EWIMP at 1$\sigma$ level from the electroweak precision measurement ($W$ boson mass measurement) at the present LHC with the experimental uncertainty being 15\,MeV, which is shown as a green solid line in both panels. Moreover, we also show in both panels the future prospect of the EWIMP detection at the 250\,GeV international linear collider (ILC) as a black dotted line assuming the integrated luminosity of 2\,ab$^{-1}$ and the beam polarizations of ($P_- = 80\%$ and $P_+: = 30\%$). This prospect is obtained from the precision measurement of the dilepton process, $e^- e^+ \to \mu^- \mu^+$, with the systematic uncertainty being 0.2\%\,\cite{Harigaya:2015yaa}. It can be seen that HL-LHC and ILC play a complementary role to search for the EWIMP; the HL-LHC has a good sensitivity for heavier EWIMPs, while the ILC has for lighter ones.

\section{Conclusion and discussion}
\label{sec: conclusion}
We have discussed in this paper the possibility of detecting EWIMP through the precision measurement of the dilepton (dielectron and dimuon) channels at the LHC. The EWIMP affects the lepton distribution of the channels through radiative corrections at ${\cal O}(1)$\% level. Since detecting such a small deviation from the SM prediction is not trivial, we have considered two different methods to analyze data. It then turned out that both gives almost the same result. Moreover, the dilepton channels are comparable to and has a potential to be better than that of the mono-$X$ search with large energy to detect the EWIMP.

Though we focus mainly on non-colored EWIMPs which are an (almost) electroweak gauge eigenstate and has a tiny decay width, the method developed here is applicable to more generic EWIMPs. For instance, a neutralino or a chargino, which is described by a mixture of different electroweak gauge eigenstates, also provide the same effects on the Drell-Yan process. Another prominent example in the SM would be a top quark.

In order to make our analysis to be more accurate, we should take the following two issues into account: First, the correlation of systematic uncertainties among the bins of the lepton invariant mass should be included, which requires more detailed information about the estimation of the SM background. Next, the radiative correction from the EWIMP at around the threshold, namely $\hat{s} \sim 4m^2$, receives a further correction by the so-called the threshold singularity when the electroweak quantum number of the EWIMP is large. Since we propose a novel idea to detect the EWIMP through the threshold observation, inclusion of these two issues is beyond our scope, and we leave it for a future work.

There are other interesting channels to detect the EWIMP at the LHC. For instance, observing the transverse mass distribution of a lepton and a missing energy (a neutrino) from the Drell-Yan process ($s-$channel exchange of the $W$ boson) is also sensitive and expected to give a similar constraint on the EWIMP. An advantage of this channel is that its cross section is larger than those of the dilepton channels that we have developed in this paper.

Another interesting aspect of the threshold observation is the measurement of the EWIMP nature rather than the discovery. As seen in Fig.\,\ref{fig: example}, the EWIMP correction depends strongly on its spin and electroweak quantum number. It would then be possible to pin down the EWIMP nature by measuring the Drell-Yan process precisely at hadron colliders.

\section*{Acknowledgments}
S. Shirai thanks M. Endo and S. Mishima for useful discussion. This work is supported by Grant-in-Aid for Scientific Research from the Ministry of Education, Culture, Sports, Science, and Technology (MEXT), Japan, No. 17H02878 (S. M. and S. S.), 16H02176 (S. M. and M. T.), 26104009 (S. M.), 17H05399 (M. T.), 16H03991 (M. T.), and by World Premier International Research Center Initiative (WPI), MEXT, Japan.

\bibliographystyle{aps}
\bibliography{ref}

\begin{thebibliography}{64}%
\makeatletter
\providecommand \@ifxundefined [1]{%
 \@ifx{#1\undefined}
}%
\providecommand \@ifnum [1]{%
 \ifnum #1\expandafter \@firstoftwo
 \else \expandafter \@secondoftwo
 \fi
}%
\providecommand \@ifx [1]{%
 \ifx #1\expandafter \@firstoftwo
 \else \expandafter \@secondoftwo
 \fi
}%
\providecommand \natexlab [1]{#1}%
\providecommand \enquote  [1]{``#1''}%
\providecommand \bibnamefont  [1]{#1}%
\providecommand \bibfnamefont [1]{#1}%
\providecommand \citenamefont [1]{#1}%
\providecommand \href@noop [0]{\@secondoftwo}%
\providecommand \href [0]{\begingroup \@sanitize@url \@href}%
\providecommand \@href[1]{\@@startlink{#1}\@@href}%
\providecommand \@@href[1]{\endgroup#1\@@endlink}%
\providecommand \@sanitize@url [0]{\catcode `\\12\catcode `\$12\catcode
  `\&12\catcode `\#12\catcode `\^12\catcode `\_12\catcode `\%12\relax}%
\providecommand \@@startlink[1]{}%
\providecommand \@@endlink[0]{}%
\providecommand \url  [0]{\begingroup\@sanitize@url \@url }%
\providecommand \@url [1]{\endgroup\@href {#1}{\urlprefix }}%
\providecommand \urlprefix  [0]{URL }%
\providecommand \Eprint [0]{\href }%
\providecommand \doibase [0]{http://dx.doi.org/}%
\providecommand \selectlanguage [0]{\@gobble}%
\providecommand \bibinfo  [0]{\@secondoftwo}%
\providecommand \bibfield  [0]{\@secondoftwo}%
\providecommand \translation [1]{[#1]}%
\providecommand \BibitemOpen [0]{}%
\providecommand \bibitemStop [0]{}%
\providecommand \bibitemNoStop [0]{.\EOS\space}%
\providecommand \EOS [0]{\spacefactor3000\relax}%
\providecommand \BibitemShut  [1]{\csname bibitem#1\endcsname}%
\let\auto@bib@innerbib\@empty
\bibitem [{\citenamefont {Hisano}\ \emph {et~al.}(2004)\citenamefont {Hisano},
  \citenamefont {Matsumoto},\ and\ \citenamefont {Nojiri}}]{Hisano:2003ec}%
  \BibitemOpen
  \bibfield  {author} {\bibinfo {author} {\bibfnamefont {J.}~\bibnamefont
  {Hisano}}, \bibinfo {author} {\bibfnamefont {S.}~\bibnamefont {Matsumoto}}, \
  and\ \bibinfo {author} {\bibfnamefont {M.~M.}\ \bibnamefont {Nojiri}},\
  }\href {\doibase 10.1103/PhysRevLett.92.031303} {\bibfield  {journal}
  {\bibinfo  {journal} {Phys. Rev. Lett.}\ }\textbf {\bibinfo {volume} {92}},\
  \bibinfo {pages} {031303} (\bibinfo {year} {2004})},\ \Eprint
  {http://arxiv.org/abs/hep-ph/0307216}{arXiv:hep-ph/0307216
  [hep-ph]}\BibitemShut {NoStop}%
\bibitem [{\citenamefont {Hisano}\ \emph {et~al.}(2005)\citenamefont {Hisano},
  \citenamefont {Matsumoto}, \citenamefont {Nojiri},\ and\ \citenamefont
  {Saito}}]{Hisano:2004ds}%
  \BibitemOpen
  \bibfield  {author} {\bibinfo {author} {\bibfnamefont {J.}~\bibnamefont
  {Hisano}}, \bibinfo {author} {\bibfnamefont {S.}~\bibnamefont {Matsumoto}},
  \bibinfo {author} {\bibfnamefont {M.~M.}\ \bibnamefont {Nojiri}}, \ and\
  \bibinfo {author} {\bibfnamefont {O.}~\bibnamefont {Saito}},\ }\href
  {\doibase 10.1103/PhysRevD.71.063528} {\bibfield  {journal} {\bibinfo
  {journal} {Phys. Rev.}\ }\textbf {\bibinfo {volume} {D71}},\ \bibinfo {pages}
  {063528} (\bibinfo {year} {2005})},\ \Eprint
  {http://arxiv.org/abs/hep-ph/0412403}{arXiv:hep-ph/0412403
  [hep-ph]}\BibitemShut {NoStop}%
\bibitem [{\citenamefont {Hisano}\ \emph {et~al.}(2006)\citenamefont {Hisano},
  \citenamefont {Matsumoto}, \citenamefont {Saito},\ and\ \citenamefont
  {Senami}}]{Hisano:2005ec}%
  \BibitemOpen
  \bibfield  {author} {\bibinfo {author} {\bibfnamefont {J.}~\bibnamefont
  {Hisano}}, \bibinfo {author} {\bibfnamefont {S.}~\bibnamefont {Matsumoto}},
  \bibinfo {author} {\bibfnamefont {O.}~\bibnamefont {Saito}}, \ and\ \bibinfo
  {author} {\bibfnamefont {M.}~\bibnamefont {Senami}},\ }\href {\doibase
  10.1103/PhysRevD.73.055004} {\bibfield  {journal} {\bibinfo  {journal} {Phys.
  Rev.}\ }\textbf {\bibinfo {volume} {D73}},\ \bibinfo {pages} {055004}
  (\bibinfo {year} {2006})},\ \Eprint
  {http://arxiv.org/abs/hep-ph/0511118}{arXiv:hep-ph/0511118
  [hep-ph]}\BibitemShut {NoStop}%
\bibitem [{\citenamefont {Hisano}\ \emph
  {et~al.}(2010{\natexlab{a}})\citenamefont {Hisano}, \citenamefont
  {Ishiwata},\ and\ \citenamefont {Nagata}}]{Hisano:2010fy}%
  \BibitemOpen
  \bibfield  {author} {\bibinfo {author} {\bibfnamefont {J.}~\bibnamefont
  {Hisano}}, \bibinfo {author} {\bibfnamefont {K.}~\bibnamefont {Ishiwata}}, \
  and\ \bibinfo {author} {\bibfnamefont {N.}~\bibnamefont {Nagata}},\ }\href
  {\doibase 10.1016/j.physletb.2010.05.047} {\bibfield  {journal} {\bibinfo
  {journal} {Phys. Lett.}\ }\textbf {\bibinfo {volume} {B690}},\ \bibinfo
  {pages} {311} (\bibinfo {year} {2010}{\natexlab{a}})},\ \Eprint
  {http://arxiv.org/abs/1004.4090}{arXiv:1004.4090 [hep-ph]}\BibitemShut
  {NoStop}%
\bibitem [{\citenamefont {Hisano}\ \emph
  {et~al.}(2010{\natexlab{b}})\citenamefont {Hisano}, \citenamefont
  {Ishiwata},\ and\ \citenamefont {Nagata}}]{Hisano:2010ct}%
  \BibitemOpen
  \bibfield  {author} {\bibinfo {author} {\bibfnamefont {J.}~\bibnamefont
  {Hisano}}, \bibinfo {author} {\bibfnamefont {K.}~\bibnamefont {Ishiwata}}, \
  and\ \bibinfo {author} {\bibfnamefont {N.}~\bibnamefont {Nagata}},\ }\href
  {\doibase 10.1103/PhysRevD.82.115007} {\bibfield  {journal} {\bibinfo
  {journal} {Phys. Rev.}\ }\textbf {\bibinfo {volume} {D82}},\ \bibinfo {pages}
  {115007} (\bibinfo {year} {2010}{\natexlab{b}})},\ \Eprint
  {http://arxiv.org/abs/1007.2601}{arXiv:1007.2601 [hep-ph]}\BibitemShut
  {NoStop}%
\bibitem [{\citenamefont {Hisano}\ \emph {et~al.}(2013)\citenamefont {Hisano},
  \citenamefont {Ishiwata},\ and\ \citenamefont {Nagata}}]{Hisano:2012wm}%
  \BibitemOpen
  \bibfield  {author} {\bibinfo {author} {\bibfnamefont {J.}~\bibnamefont
  {Hisano}}, \bibinfo {author} {\bibfnamefont {K.}~\bibnamefont {Ishiwata}}, \
  and\ \bibinfo {author} {\bibfnamefont {N.}~\bibnamefont {Nagata}},\ }\href
  {\doibase 10.1103/PhysRevD.87.035020} {\bibfield  {journal} {\bibinfo
  {journal} {Phys. Rev.}\ }\textbf {\bibinfo {volume} {D87}},\ \bibinfo {pages}
  {035020} (\bibinfo {year} {2013})},\ \Eprint
  {http://arxiv.org/abs/1210.5985}{arXiv:1210.5985 [hep-ph]}\BibitemShut
  {NoStop}%
\bibitem [{\citenamefont {Hisano}\ \emph
  {et~al.}(2015{\natexlab{a}})\citenamefont {Hisano}, \citenamefont
  {Ishiwata},\ and\ \citenamefont {Nagata}}]{Hisano:2015rsa}%
  \BibitemOpen
  \bibfield  {author} {\bibinfo {author} {\bibfnamefont {J.}~\bibnamefont
  {Hisano}}, \bibinfo {author} {\bibfnamefont {K.}~\bibnamefont {Ishiwata}}, \
  and\ \bibinfo {author} {\bibfnamefont {N.}~\bibnamefont {Nagata}},\ }\href
  {\doibase 10.1007/JHEP06(2015)097} {\bibfield  {journal} {\bibinfo  {journal}
  {JHEP}\ }\textbf {\bibinfo {volume} {06}},\ \bibinfo {pages} {097} (\bibinfo
  {year} {2015}{\natexlab{a}})},\ \Eprint
  {http://arxiv.org/abs/1504.00915}{arXiv:1504.00915 [hep-ph]}\BibitemShut
  {NoStop}%
\bibitem [{\citenamefont {Randall}\ and\ \citenamefont
  {Sundrum}(1999)}]{Randall:1998uk}%
  \BibitemOpen
  \bibfield  {author} {\bibinfo {author} {\bibfnamefont {L.}~\bibnamefont
  {Randall}}\ and\ \bibinfo {author} {\bibfnamefont {R.}~\bibnamefont
  {Sundrum}},\ }\href {\doibase 10.1016/S0550-3213(99)00359-4} {\bibfield
  {journal} {\bibinfo  {journal} {Nucl. Phys.}\ }\textbf {\bibinfo {volume}
  {B557}},\ \bibinfo {pages} {79} (\bibinfo {year} {1999})},\ \Eprint
  {http://arxiv.org/abs/hep-th/9810155}{arXiv:hep-th/9810155
  [hep-th]}\BibitemShut {NoStop}%
\bibitem [{\citenamefont {Giudice}\ \emph {et~al.}(1998)\citenamefont
  {Giudice}, \citenamefont {Luty}, \citenamefont {Murayama},\ and\
  \citenamefont {Rattazzi}}]{Giudice:1998xp}%
  \BibitemOpen
  \bibfield  {author} {\bibinfo {author} {\bibfnamefont {G.~F.}\ \bibnamefont
  {Giudice}}, \bibinfo {author} {\bibfnamefont {M.~A.}\ \bibnamefont {Luty}},
  \bibinfo {author} {\bibfnamefont {H.}~\bibnamefont {Murayama}}, \ and\
  \bibinfo {author} {\bibfnamefont {R.}~\bibnamefont {Rattazzi}},\ }\href
  {\doibase 10.1088/1126-6708/1998/12/027} {\bibfield  {journal} {\bibinfo
  {journal} {JHEP}\ }\textbf {\bibinfo {volume} {12}},\ \bibinfo {pages} {027}
  (\bibinfo {year} {1998})},\ \Eprint
  {http://arxiv.org/abs/hep-ph/9810442}{arXiv:hep-ph/9810442
  [hep-ph]}\BibitemShut {NoStop}%
\bibitem [{\citenamefont {Wells}(2003)}]{Wells:2003tf}%
  \BibitemOpen
  \bibfield  {author} {\bibinfo {author} {\bibfnamefont {J.~D.}\ \bibnamefont
  {Wells}},\ }in\ \href@noop {} {\emph {\bibinfo {booktitle} {{11th
  International Conference on Supersymmetry and the Unification of Fundamental
  Interactions (SUSY 2003) Tucson, Arizona, June 5-10, 2003}}}}\ (\bibinfo
  {year} {2003})\ \Eprint
  {http://arxiv.org/abs/hep-ph/0306127}{arXiv:hep-ph/0306127
  [hep-ph]}\BibitemShut {NoStop}%
\bibitem [{\citenamefont {Wells}(2005)}]{Wells:2004di}%
  \BibitemOpen
  \bibfield  {author} {\bibinfo {author} {\bibfnamefont {J.~D.}\ \bibnamefont
  {Wells}},\ }\href {\doibase 10.1103/PhysRevD.71.015013} {\bibfield  {journal}
  {\bibinfo  {journal} {Phys. Rev.}\ }\textbf {\bibinfo {volume} {D71}},\
  \bibinfo {pages} {015013} (\bibinfo {year} {2005})},\ \Eprint
  {http://arxiv.org/abs/hep-ph/0411041}{arXiv:hep-ph/0411041
  [hep-ph]}\BibitemShut {NoStop}%
\bibitem [{\citenamefont {Arkani-Hamed}\ and\ \citenamefont
  {Dimopoulos}(2005)}]{ArkaniHamed:2004fb}%
  \BibitemOpen
  \bibfield  {author} {\bibinfo {author} {\bibfnamefont {N.}~\bibnamefont
  {Arkani-Hamed}}\ and\ \bibinfo {author} {\bibfnamefont {S.}~\bibnamefont
  {Dimopoulos}},\ }\href {\doibase 10.1088/1126-6708/2005/06/073} {\bibfield
  {journal} {\bibinfo  {journal} {JHEP}\ }\textbf {\bibinfo {volume} {06}},\
  \bibinfo {pages} {073} (\bibinfo {year} {2005})},\ \Eprint
  {http://arxiv.org/abs/hep-th/0405159}{arXiv:hep-th/0405159
  [hep-th]}\BibitemShut {NoStop}%
\bibitem [{\citenamefont {Giudice}\ and\ \citenamefont
  {Romanino}(2004)}]{Giudice:2004tc}%
  \BibitemOpen
  \bibfield  {author} {\bibinfo {author} {\bibfnamefont {G.~F.}\ \bibnamefont
  {Giudice}}\ and\ \bibinfo {author} {\bibfnamefont {A.}~\bibnamefont
  {Romanino}},\ }\href {\doibase 10.1016/j.nuclphysb.2004.11.048,
  10.1016/j.nuclphysb.2004.08.001} {\bibfield  {journal} {\bibinfo  {journal}
  {Nucl. Phys.}\ }\textbf {\bibinfo {volume} {B699}},\ \bibinfo {pages} {65}
  (\bibinfo {year} {2004})},\ \bibinfo {note} {[Erratum: Nucl.
  Phys.B706,487(2005)]},\ \Eprint
  {http://arxiv.org/abs/hep-ph/0406088}{arXiv:hep-ph/0406088
  [hep-ph]}\BibitemShut {NoStop}%
\bibitem [{\citenamefont {Arkani-Hamed}\ \emph
  {et~al.}(2005{\natexlab{a}})\citenamefont {Arkani-Hamed}, \citenamefont
  {Dimopoulos}, \citenamefont {Giudice},\ and\ \citenamefont
  {Romanino}}]{ArkaniHamed:2004yi}%
  \BibitemOpen
  \bibfield  {author} {\bibinfo {author} {\bibfnamefont {N.}~\bibnamefont
  {Arkani-Hamed}}, \bibinfo {author} {\bibfnamefont {S.}~\bibnamefont
  {Dimopoulos}}, \bibinfo {author} {\bibfnamefont {G.~F.}\ \bibnamefont
  {Giudice}}, \ and\ \bibinfo {author} {\bibfnamefont {A.}~\bibnamefont
  {Romanino}},\ }\href {\doibase 10.1016/j.nuclphysb.2004.12.026} {\bibfield
  {journal} {\bibinfo  {journal} {Nucl. Phys.}\ }\textbf {\bibinfo {volume}
  {B709}},\ \bibinfo {pages} {3} (\bibinfo {year} {2005}{\natexlab{a}})},\
  \Eprint {http://arxiv.org/abs/hep-ph/0409232}{arXiv:hep-ph/0409232
  [hep-ph]}\BibitemShut {NoStop}%
\bibitem [{\citenamefont {Arkani-Hamed}\ \emph
  {et~al.}(2005{\natexlab{b}})\citenamefont {Arkani-Hamed}, \citenamefont
  {Dimopoulos},\ and\ \citenamefont {Kachru}}]{ArkaniHamed:2005yv}%
  \BibitemOpen
  \bibfield  {author} {\bibinfo {author} {\bibfnamefont {N.}~\bibnamefont
  {Arkani-Hamed}}, \bibinfo {author} {\bibfnamefont {S.}~\bibnamefont
  {Dimopoulos}}, \ and\ \bibinfo {author} {\bibfnamefont {S.}~\bibnamefont
  {Kachru}},\ }\href@noop {} {\  (\bibinfo {year} {2005}{\natexlab{b}})},\
  \Eprint {http://arxiv.org/abs/hep-th/0501082}{arXiv:hep-th/0501082
  [hep-th]}\BibitemShut {NoStop}%
\bibitem [{\citenamefont {Aad}\ \emph {et~al.}(2012)\citenamefont {Aad} \emph
  {et~al.}}]{Aad:2012tfa}%
  \BibitemOpen
  \bibfield  {author} {\bibinfo {author} {\bibfnamefont {G.}~\bibnamefont
  {Aad}} \emph {et~al.} (\bibinfo {collaboration} {ATLAS}),\ }\href {\doibase
  10.1016/j.physletb.2012.08.020} {\bibfield  {journal} {\bibinfo  {journal}
  {Phys. Lett.}\ }\textbf {\bibinfo {volume} {B716}},\ \bibinfo {pages} {1}
  (\bibinfo {year} {2012})},\ \Eprint
  {http://arxiv.org/abs/1207.7214}{arXiv:1207.7214 [hep-ex]}\BibitemShut
  {NoStop}%
\bibitem [{\citenamefont {Chatrchyan}\ \emph {et~al.}(2012)\citenamefont
  {Chatrchyan} \emph {et~al.}}]{Chatrchyan:2012ufa}%
  \BibitemOpen
  \bibfield  {author} {\bibinfo {author} {\bibfnamefont {S.}~\bibnamefont
  {Chatrchyan}} \emph {et~al.} (\bibinfo {collaboration} {CMS}),\ }\href
  {\doibase 10.1016/j.physletb.2012.08.021} {\bibfield  {journal} {\bibinfo
  {journal} {Phys. Lett.}\ }\textbf {\bibinfo {volume} {B716}},\ \bibinfo
  {pages} {30} (\bibinfo {year} {2012})},\ \Eprint
  {http://arxiv.org/abs/1207.7235}{arXiv:1207.7235 [hep-ex]}\BibitemShut
  {NoStop}%
\bibitem [{\citenamefont {Hall}\ and\ \citenamefont
  {Nomura}(2012)}]{Hall:2011jd}%
  \BibitemOpen
  \bibfield  {author} {\bibinfo {author} {\bibfnamefont {L.~J.}\ \bibnamefont
  {Hall}}\ and\ \bibinfo {author} {\bibfnamefont {Y.}~\bibnamefont {Nomura}},\
  }\href {\doibase 10.1007/JHEP01(2012)082} {\bibfield  {journal} {\bibinfo
  {journal} {JHEP}\ }\textbf {\bibinfo {volume} {01}},\ \bibinfo {pages} {082}
  (\bibinfo {year} {2012})},\ \Eprint
  {http://arxiv.org/abs/1111.4519}{arXiv:1111.4519 [hep-ph]}\BibitemShut
  {NoStop}%
\bibitem [{\citenamefont {Hall}\ \emph {et~al.}(2013)\citenamefont {Hall},
  \citenamefont {Nomura},\ and\ \citenamefont {Shirai}}]{Hall:2012zp}%
  \BibitemOpen
  \bibfield  {author} {\bibinfo {author} {\bibfnamefont {L.~J.}\ \bibnamefont
  {Hall}}, \bibinfo {author} {\bibfnamefont {Y.}~\bibnamefont {Nomura}}, \ and\
  \bibinfo {author} {\bibfnamefont {S.}~\bibnamefont {Shirai}},\ }\href
  {\doibase 10.1007/JHEP01(2013)036} {\bibfield  {journal} {\bibinfo  {journal}
  {JHEP}\ }\textbf {\bibinfo {volume} {01}},\ \bibinfo {pages} {036} (\bibinfo
  {year} {2013})},\ \Eprint {http://arxiv.org/abs/1210.2395}{arXiv:1210.2395
  [hep-ph]}\BibitemShut {NoStop}%
\bibitem [{\citenamefont {Nomura}\ and\ \citenamefont
  {Shirai}(2014)}]{Nomura:2014asa}%
  \BibitemOpen
  \bibfield  {author} {\bibinfo {author} {\bibfnamefont {Y.}~\bibnamefont
  {Nomura}}\ and\ \bibinfo {author} {\bibfnamefont {S.}~\bibnamefont
  {Shirai}},\ }\href {\doibase 10.1103/PhysRevLett.113.111801} {\bibfield
  {journal} {\bibinfo  {journal} {Phys. Rev. Lett.}\ }\textbf {\bibinfo
  {volume} {113}},\ \bibinfo {pages} {111801} (\bibinfo {year} {2014})},\
  \Eprint {http://arxiv.org/abs/1407.3785}{arXiv:1407.3785
  [hep-ph]}\BibitemShut {NoStop}%
\bibitem [{\citenamefont {Ibe}\ and\ \citenamefont
  {Yanagida}(2012)}]{Ibe:2011aa}%
  \BibitemOpen
  \bibfield  {author} {\bibinfo {author} {\bibfnamefont {M.}~\bibnamefont
  {Ibe}}\ and\ \bibinfo {author} {\bibfnamefont {T.~T.}\ \bibnamefont
  {Yanagida}},\ }\href {\doibase 10.1016/j.physletb.2012.02.034} {\bibfield
  {journal} {\bibinfo  {journal} {Phys. Lett.}\ }\textbf {\bibinfo {volume}
  {B709}},\ \bibinfo {pages} {374} (\bibinfo {year} {2012})},\ \Eprint
  {http://arxiv.org/abs/1112.2462}{arXiv:1112.2462 [hep-ph]}\BibitemShut
  {NoStop}%
\bibitem [{\citenamefont {Ibe}\ \emph {et~al.}(2012)\citenamefont {Ibe},
  \citenamefont {Matsumoto},\ and\ \citenamefont {Yanagida}}]{Ibe:2012hu}%
  \BibitemOpen
  \bibfield  {author} {\bibinfo {author} {\bibfnamefont {M.}~\bibnamefont
  {Ibe}}, \bibinfo {author} {\bibfnamefont {S.}~\bibnamefont {Matsumoto}}, \
  and\ \bibinfo {author} {\bibfnamefont {T.~T.}\ \bibnamefont {Yanagida}},\
  }\href {\doibase 10.1103/PhysRevD.85.095011} {\bibfield  {journal} {\bibinfo
  {journal} {Phys. Rev.}\ }\textbf {\bibinfo {volume} {D85}},\ \bibinfo {pages}
  {095011} (\bibinfo {year} {2012})},\ \Eprint
  {http://arxiv.org/abs/1202.2253}{arXiv:1202.2253 [hep-ph]}\BibitemShut
  {NoStop}%
\bibitem [{\citenamefont {Arvanitaki}\ \emph {et~al.}(2013)\citenamefont
  {Arvanitaki}, \citenamefont {Craig}, \citenamefont {Dimopoulos},\ and\
  \citenamefont {Villadoro}}]{Arvanitaki:2012ps}%
  \BibitemOpen
  \bibfield  {author} {\bibinfo {author} {\bibfnamefont {A.}~\bibnamefont
  {Arvanitaki}}, \bibinfo {author} {\bibfnamefont {N.}~\bibnamefont {Craig}},
  \bibinfo {author} {\bibfnamefont {S.}~\bibnamefont {Dimopoulos}}, \ and\
  \bibinfo {author} {\bibfnamefont {G.}~\bibnamefont {Villadoro}},\ }\href
  {\doibase 10.1007/JHEP02(2013)126} {\bibfield  {journal} {\bibinfo  {journal}
  {JHEP}\ }\textbf {\bibinfo {volume} {02}},\ \bibinfo {pages} {126} (\bibinfo
  {year} {2013})},\ \Eprint {http://arxiv.org/abs/1210.0555}{arXiv:1210.0555
  [hep-ph]}\BibitemShut {NoStop}%
\bibitem [{\citenamefont {Arkani-Hamed}\ \emph {et~al.}(2012)\citenamefont
  {Arkani-Hamed}, \citenamefont {Gupta}, \citenamefont {Kaplan}, \citenamefont
  {Weiner},\ and\ \citenamefont {Zorawski}}]{ArkaniHamed:2012gw}%
  \BibitemOpen
  \bibfield  {author} {\bibinfo {author} {\bibfnamefont {N.}~\bibnamefont
  {Arkani-Hamed}}, \bibinfo {author} {\bibfnamefont {A.}~\bibnamefont {Gupta}},
  \bibinfo {author} {\bibfnamefont {D.~E.}\ \bibnamefont {Kaplan}}, \bibinfo
  {author} {\bibfnamefont {N.}~\bibnamefont {Weiner}}, \ and\ \bibinfo {author}
  {\bibfnamefont {T.}~\bibnamefont {Zorawski}},\ }\href@noop {} {\  (\bibinfo
  {year} {2012})},\ \Eprint {http://arxiv.org/abs/1212.6971}{arXiv:1212.6971
  [hep-ph]}\BibitemShut {NoStop}%
\bibitem [{\citenamefont {Cohen}\ \emph {et~al.}(2013)\citenamefont {Cohen},
  \citenamefont {Lisanti}, \citenamefont {Pierce},\ and\ \citenamefont
  {Slatyer}}]{Cohen:2013ama}%
  \BibitemOpen
  \bibfield  {author} {\bibinfo {author} {\bibfnamefont {T.}~\bibnamefont
  {Cohen}}, \bibinfo {author} {\bibfnamefont {M.}~\bibnamefont {Lisanti}},
  \bibinfo {author} {\bibfnamefont {A.}~\bibnamefont {Pierce}}, \ and\ \bibinfo
  {author} {\bibfnamefont {T.~R.}\ \bibnamefont {Slatyer}},\ }\href {\doibase
  10.1088/1475-7516/2013/10/061} {\bibfield  {journal} {\bibinfo  {journal}
  {JCAP}\ }\textbf {\bibinfo {volume} {1310}},\ \bibinfo {pages} {061}
  (\bibinfo {year} {2013})},\ \Eprint
  {http://arxiv.org/abs/1307.4082}{arXiv:1307.4082 [hep-ph]}\BibitemShut
  {NoStop}%
\bibitem [{\citenamefont {Fan}\ and\ \citenamefont
  {Reece}(2013)}]{Fan:2013faa}%
  \BibitemOpen
  \bibfield  {author} {\bibinfo {author} {\bibfnamefont {J.}~\bibnamefont
  {Fan}}\ and\ \bibinfo {author} {\bibfnamefont {M.}~\bibnamefont {Reece}},\
  }\href {\doibase 10.1007/JHEP10(2013)124} {\bibfield  {journal} {\bibinfo
  {journal} {JHEP}\ }\textbf {\bibinfo {volume} {10}},\ \bibinfo {pages} {124}
  (\bibinfo {year} {2013})},\ \Eprint
  {http://arxiv.org/abs/1307.4400}{arXiv:1307.4400 [hep-ph]}\BibitemShut
  {NoStop}%
\bibitem [{\citenamefont {Bhattacherjee}\ \emph {et~al.}(2014)\citenamefont
  {Bhattacherjee}, \citenamefont {Ibe}, \citenamefont {Ichikawa}, \citenamefont
  {Matsumoto},\ and\ \citenamefont {Nishiyama}}]{Bhattacherjee:2014dya}%
  \BibitemOpen
  \bibfield  {author} {\bibinfo {author} {\bibfnamefont {B.}~\bibnamefont
  {Bhattacherjee}}, \bibinfo {author} {\bibfnamefont {M.}~\bibnamefont {Ibe}},
  \bibinfo {author} {\bibfnamefont {K.}~\bibnamefont {Ichikawa}}, \bibinfo
  {author} {\bibfnamefont {S.}~\bibnamefont {Matsumoto}}, \ and\ \bibinfo
  {author} {\bibfnamefont {K.}~\bibnamefont {Nishiyama}},\ }\href {\doibase
  10.1007/JHEP07(2014)080} {\bibfield  {journal} {\bibinfo  {journal} {JHEP}\
  }\textbf {\bibinfo {volume} {07}},\ \bibinfo {pages} {080} (\bibinfo {year}
  {2014})},\ \Eprint {http://arxiv.org/abs/1405.4914}{arXiv:1405.4914
  [hep-ph]}\BibitemShut {NoStop}%
\bibitem [{\citenamefont {Ibe}\ \emph {et~al.}(2015)\citenamefont {Ibe},
  \citenamefont {Matsumoto}, \citenamefont {Shirai},\ and\ \citenamefont
  {Yanagida}}]{Ibe:2015tma}%
  \BibitemOpen
  \bibfield  {author} {\bibinfo {author} {\bibfnamefont {M.}~\bibnamefont
  {Ibe}}, \bibinfo {author} {\bibfnamefont {S.}~\bibnamefont {Matsumoto}},
  \bibinfo {author} {\bibfnamefont {S.}~\bibnamefont {Shirai}}, \ and\ \bibinfo
  {author} {\bibfnamefont {T.~T.}\ \bibnamefont {Yanagida}},\ }\href {\doibase
  10.1103/PhysRevD.91.111701} {\bibfield  {journal} {\bibinfo  {journal} {Phys.
  Rev.}\ }\textbf {\bibinfo {volume} {D91}},\ \bibinfo {pages} {111701}
  (\bibinfo {year} {2015})},\ \Eprint
  {http://arxiv.org/abs/1504.05554}{arXiv:1504.05554 [hep-ph]}\BibitemShut
  {NoStop}%
\bibitem [{\citenamefont {Cirelli}\ \emph {et~al.}(2006)\citenamefont
  {Cirelli}, \citenamefont {Fornengo},\ and\ \citenamefont
  {Strumia}}]{Cirelli:2005uq}%
  \BibitemOpen
  \bibfield  {author} {\bibinfo {author} {\bibfnamefont {M.}~\bibnamefont
  {Cirelli}}, \bibinfo {author} {\bibfnamefont {N.}~\bibnamefont {Fornengo}}, \
  and\ \bibinfo {author} {\bibfnamefont {A.}~\bibnamefont {Strumia}},\ }\href
  {\doibase 10.1016/j.nuclphysb.2006.07.012} {\bibfield  {journal} {\bibinfo
  {journal} {Nucl. Phys.}\ }\textbf {\bibinfo {volume} {B753}},\ \bibinfo
  {pages} {178} (\bibinfo {year} {2006})},\ \Eprint
  {http://arxiv.org/abs/hep-ph/0512090}{arXiv:hep-ph/0512090
  [hep-ph]}\BibitemShut {NoStop}%
\bibitem [{\citenamefont {Cirelli}\ \emph {et~al.}(2007)\citenamefont
  {Cirelli}, \citenamefont {Strumia},\ and\ \citenamefont
  {Tamburini}}]{Cirelli:2007xd}%
  \BibitemOpen
  \bibfield  {author} {\bibinfo {author} {\bibfnamefont {M.}~\bibnamefont
  {Cirelli}}, \bibinfo {author} {\bibfnamefont {A.}~\bibnamefont {Strumia}}, \
  and\ \bibinfo {author} {\bibfnamefont {M.}~\bibnamefont {Tamburini}},\ }\href
  {\doibase 10.1016/j.nuclphysb.2007.07.023} {\bibfield  {journal} {\bibinfo
  {journal} {Nucl. Phys.}\ }\textbf {\bibinfo {volume} {B787}},\ \bibinfo
  {pages} {152} (\bibinfo {year} {2007})},\ \Eprint
  {http://arxiv.org/abs/0706.4071}{arXiv:0706.4071 [hep-ph]}\BibitemShut
  {NoStop}%
\bibitem [{\citenamefont {Cirelli}\ and\ \citenamefont
  {Strumia}(2009)}]{Cirelli:2009uv}%
  \BibitemOpen
  \bibfield  {author} {\bibinfo {author} {\bibfnamefont {M.}~\bibnamefont
  {Cirelli}}\ and\ \bibinfo {author} {\bibfnamefont {A.}~\bibnamefont
  {Strumia}},\ }\href {\doibase 10.1088/1367-2630/11/10/105005} {\bibfield
  {journal} {\bibinfo  {journal} {New J. Phys.}\ }\textbf {\bibinfo {volume}
  {11}},\ \bibinfo {pages} {105005} (\bibinfo {year} {2009})},\ \Eprint
  {http://arxiv.org/abs/0903.3381}{arXiv:0903.3381 [hep-ph]}\BibitemShut
  {NoStop}%
\bibitem [{\citenamefont {Kitano}\ and\ \citenamefont
  {Nomura}(2005)}]{Kitano:2005wc}%
  \BibitemOpen
  \bibfield  {author} {\bibinfo {author} {\bibfnamefont {R.}~\bibnamefont
  {Kitano}}\ and\ \bibinfo {author} {\bibfnamefont {Y.}~\bibnamefont
  {Nomura}},\ }\href {\doibase 10.1016/j.physletb.2005.10.003} {\bibfield
  {journal} {\bibinfo  {journal} {Phys. Lett.}\ }\textbf {\bibinfo {volume}
  {B631}},\ \bibinfo {pages} {58} (\bibinfo {year} {2005})},\ \Eprint
  {http://arxiv.org/abs/hep-ph/0509039}{arXiv:hep-ph/0509039
  [hep-ph]}\BibitemShut {NoStop}%
\bibitem [{\citenamefont {Baer}\ \emph {et~al.}(2012)\citenamefont {Baer},
  \citenamefont {Barger}, \citenamefont {Huang}, \citenamefont {Mustafayev},\
  and\ \citenamefont {Tata}}]{Baer:2012up}%
  \BibitemOpen
  \bibfield  {author} {\bibinfo {author} {\bibfnamefont {H.}~\bibnamefont
  {Baer}}, \bibinfo {author} {\bibfnamefont {V.}~\bibnamefont {Barger}},
  \bibinfo {author} {\bibfnamefont {P.}~\bibnamefont {Huang}}, \bibinfo
  {author} {\bibfnamefont {A.}~\bibnamefont {Mustafayev}}, \ and\ \bibinfo
  {author} {\bibfnamefont {X.}~\bibnamefont {Tata}},\ }\href {\doibase
  10.1103/PhysRevLett.109.161802} {\bibfield  {journal} {\bibinfo  {journal}
  {Phys. Rev. Lett.}\ }\textbf {\bibinfo {volume} {109}},\ \bibinfo {pages}
  {161802} (\bibinfo {year} {2012})},\ \Eprint
  {http://arxiv.org/abs/1207.3343}{arXiv:1207.3343 [hep-ph]}\BibitemShut
  {NoStop}%
\bibitem [{\citenamefont {Baer}\ \emph {et~al.}(2013)\citenamefont {Baer},
  \citenamefont {Barger},\ and\ \citenamefont {Mickelson}}]{Baer:2013gva}%
  \BibitemOpen
  \bibfield  {author} {\bibinfo {author} {\bibfnamefont {H.}~\bibnamefont
  {Baer}}, \bibinfo {author} {\bibfnamefont {V.}~\bibnamefont {Barger}}, \ and\
  \bibinfo {author} {\bibfnamefont {D.}~\bibnamefont {Mickelson}},\ }\href
  {\doibase 10.1103/PhysRevD.88.095013} {\bibfield  {journal} {\bibinfo
  {journal} {Phys. Rev.}\ }\textbf {\bibinfo {volume} {D88}},\ \bibinfo {pages}
  {095013} (\bibinfo {year} {2013})},\ \Eprint
  {http://arxiv.org/abs/1309.2984}{arXiv:1309.2984 [hep-ph]}\BibitemShut
  {NoStop}%
\bibitem [{\citenamefont {Gori}\ \emph {et~al.}(2013)\citenamefont {Gori},
  \citenamefont {Jung},\ and\ \citenamefont {Wang}}]{Gori:2013ala}%
  \BibitemOpen
  \bibfield  {author} {\bibinfo {author} {\bibfnamefont {S.}~\bibnamefont
  {Gori}}, \bibinfo {author} {\bibfnamefont {S.}~\bibnamefont {Jung}}, \ and\
  \bibinfo {author} {\bibfnamefont {L.-T.}\ \bibnamefont {Wang}},\ }\href
  {\doibase 10.1007/JHEP10(2013)191} {\bibfield  {journal} {\bibinfo  {journal}
  {JHEP}\ }\textbf {\bibinfo {volume} {10}},\ \bibinfo {pages} {191} (\bibinfo
  {year} {2013})},\ \Eprint {http://arxiv.org/abs/1307.5952}{arXiv:1307.5952
  [hep-ph]}\BibitemShut {NoStop}%
\bibitem [{\citenamefont {Han}\ \emph {et~al.}(2014{\natexlab{a}})\citenamefont
  {Han}, \citenamefont {Kobakhidze}, \citenamefont {Liu}, \citenamefont
  {Saavedra}, \citenamefont {Wu},\ and\ \citenamefont {Yang}}]{Han:2013usa}%
  \BibitemOpen
  \bibfield  {author} {\bibinfo {author} {\bibfnamefont {C.}~\bibnamefont
  {Han}}, \bibinfo {author} {\bibfnamefont {A.}~\bibnamefont {Kobakhidze}},
  \bibinfo {author} {\bibfnamefont {N.}~\bibnamefont {Liu}}, \bibinfo {author}
  {\bibfnamefont {A.}~\bibnamefont {Saavedra}}, \bibinfo {author}
  {\bibfnamefont {L.}~\bibnamefont {Wu}}, \ and\ \bibinfo {author}
  {\bibfnamefont {J.~M.}\ \bibnamefont {Yang}},\ }\href {\doibase
  10.1007/JHEP02(2014)049} {\bibfield  {journal} {\bibinfo  {journal} {JHEP}\
  }\textbf {\bibinfo {volume} {02}},\ \bibinfo {pages} {049} (\bibinfo {year}
  {2014}{\natexlab{a}})},\ \Eprint
  {http://arxiv.org/abs/1310.4274}{arXiv:1310.4274 [hep-ph]}\BibitemShut
  {NoStop}%
\bibitem [{\citenamefont {Han}\ \emph {et~al.}(2014{\natexlab{b}})\citenamefont
  {Han}, \citenamefont {Kribs}, \citenamefont {Martin},\ and\ \citenamefont
  {Menon}}]{Han:2014kaa}%
  \BibitemOpen
  \bibfield  {author} {\bibinfo {author} {\bibfnamefont {Z.}~\bibnamefont
  {Han}}, \bibinfo {author} {\bibfnamefont {G.~D.}\ \bibnamefont {Kribs}},
  \bibinfo {author} {\bibfnamefont {A.}~\bibnamefont {Martin}}, \ and\ \bibinfo
  {author} {\bibfnamefont {A.}~\bibnamefont {Menon}},\ }\href {\doibase
  10.1103/PhysRevD.89.075007} {\bibfield  {journal} {\bibinfo  {journal} {Phys.
  Rev.}\ }\textbf {\bibinfo {volume} {D89}},\ \bibinfo {pages} {075007}
  (\bibinfo {year} {2014}{\natexlab{b}})},\ \Eprint
  {http://arxiv.org/abs/1401.1235}{arXiv:1401.1235 [hep-ph]}\BibitemShut
  {NoStop}%
\bibitem [{\citenamefont {Bramante}\ \emph {et~al.}(2014)\citenamefont
  {Bramante}, \citenamefont {Delgado}, \citenamefont {Elahi}, \citenamefont
  {Martin},\ and\ \citenamefont {Ostdiek}}]{Bramante:2014dza}%
  \BibitemOpen
  \bibfield  {author} {\bibinfo {author} {\bibfnamefont {J.}~\bibnamefont
  {Bramante}}, \bibinfo {author} {\bibfnamefont {A.}~\bibnamefont {Delgado}},
  \bibinfo {author} {\bibfnamefont {F.}~\bibnamefont {Elahi}}, \bibinfo
  {author} {\bibfnamefont {A.}~\bibnamefont {Martin}}, \ and\ \bibinfo {author}
  {\bibfnamefont {B.}~\bibnamefont {Ostdiek}},\ }\href {\doibase
  10.1103/PhysRevD.90.095008} {\bibfield  {journal} {\bibinfo  {journal} {Phys.
  Rev.}\ }\textbf {\bibinfo {volume} {D90}},\ \bibinfo {pages} {095008}
  (\bibinfo {year} {2014})},\ \Eprint
  {http://arxiv.org/abs/1408.6530}{arXiv:1408.6530 [hep-ph]}\BibitemShut
  {NoStop}%
\bibitem [{\citenamefont {Baer}\ \emph {et~al.}(2014)\citenamefont {Baer},
  \citenamefont {Mustafayev},\ and\ \citenamefont {Tata}}]{Baer:2014kya}%
  \BibitemOpen
  \bibfield  {author} {\bibinfo {author} {\bibfnamefont {H.}~\bibnamefont
  {Baer}}, \bibinfo {author} {\bibfnamefont {A.}~\bibnamefont {Mustafayev}}, \
  and\ \bibinfo {author} {\bibfnamefont {X.}~\bibnamefont {Tata}},\ }\href
  {\doibase 10.1103/PhysRevD.90.115007} {\bibfield  {journal} {\bibinfo
  {journal} {Phys. Rev.}\ }\textbf {\bibinfo {volume} {D90}},\ \bibinfo {pages}
  {115007} (\bibinfo {year} {2014})},\ \Eprint
  {http://arxiv.org/abs/1409.7058}{arXiv:1409.7058 [hep-ph]}\BibitemShut
  {NoStop}%
\bibitem [{\citenamefont {Bramante}\ \emph {et~al.}(2015)\citenamefont
  {Bramante}, \citenamefont {Fox}, \citenamefont {Martin}, \citenamefont
  {Ostdiek}, \citenamefont {Plehn}, \citenamefont {Schell},\ and\ \citenamefont
  {Takeuchi}}]{Bramante:2014tba}%
  \BibitemOpen
  \bibfield  {author} {\bibinfo {author} {\bibfnamefont {J.}~\bibnamefont
  {Bramante}}, \bibinfo {author} {\bibfnamefont {P.~J.}\ \bibnamefont {Fox}},
  \bibinfo {author} {\bibfnamefont {A.}~\bibnamefont {Martin}}, \bibinfo
  {author} {\bibfnamefont {B.}~\bibnamefont {Ostdiek}}, \bibinfo {author}
  {\bibfnamefont {T.}~\bibnamefont {Plehn}}, \bibinfo {author} {\bibfnamefont
  {T.}~\bibnamefont {Schell}}, \ and\ \bibinfo {author} {\bibfnamefont
  {M.}~\bibnamefont {Takeuchi}},\ }\href {\doibase 10.1103/PhysRevD.91.054015}
  {\bibfield  {journal} {\bibinfo  {journal} {Phys. Rev.}\ }\textbf {\bibinfo
  {volume} {D91}},\ \bibinfo {pages} {054015} (\bibinfo {year} {2015})},\
  \Eprint {http://arxiv.org/abs/1412.4789}{arXiv:1412.4789
  [hep-ph]}\BibitemShut {NoStop}%
\bibitem [{\citenamefont {Ismail}\ \emph {et~al.}(2016)\citenamefont {Ismail},
  \citenamefont {Izaguirre},\ and\ \citenamefont {Shuve}}]{Ismail:2016zby}%
  \BibitemOpen
  \bibfield  {author} {\bibinfo {author} {\bibfnamefont {A.}~\bibnamefont
  {Ismail}}, \bibinfo {author} {\bibfnamefont {E.}~\bibnamefont {Izaguirre}}, \
  and\ \bibinfo {author} {\bibfnamefont {B.}~\bibnamefont {Shuve}},\ }\href
  {\doibase 10.1103/PhysRevD.94.015001} {\bibfield  {journal} {\bibinfo
  {journal} {Phys. Rev.}\ }\textbf {\bibinfo {volume} {D94}},\ \bibinfo {pages}
  {015001} (\bibinfo {year} {2016})},\ \Eprint
  {http://arxiv.org/abs/1605.00658}{arXiv:1605.00658 [hep-ph]}\BibitemShut
  {NoStop}%
\bibitem [{\citenamefont {Ibe}\ \emph {et~al.}(2007)\citenamefont {Ibe},
  \citenamefont {Moroi},\ and\ \citenamefont {Yanagida}}]{Ibe:2006de}%
  \BibitemOpen
  \bibfield  {author} {\bibinfo {author} {\bibfnamefont {M.}~\bibnamefont
  {Ibe}}, \bibinfo {author} {\bibfnamefont {T.}~\bibnamefont {Moroi}}, \ and\
  \bibinfo {author} {\bibfnamefont {T.~T.}\ \bibnamefont {Yanagida}},\ }\href
  {\doibase 10.1016/j.physletb.2006.11.061} {\bibfield  {journal} {\bibinfo
  {journal} {Phys. Lett.}\ }\textbf {\bibinfo {volume} {B644}},\ \bibinfo
  {pages} {355} (\bibinfo {year} {2007})},\ \Eprint
  {http://arxiv.org/abs/hep-ph/0610277}{arXiv:hep-ph/0610277
  [hep-ph]}\BibitemShut {NoStop}%
\bibitem [{\citenamefont {Buckley}\ \emph {et~al.}(2011)\citenamefont
  {Buckley}, \citenamefont {Randall},\ and\ \citenamefont
  {Shuve}}]{Buckley:2009kv}%
  \BibitemOpen
  \bibfield  {author} {\bibinfo {author} {\bibfnamefont {M.~R.}\ \bibnamefont
  {Buckley}}, \bibinfo {author} {\bibfnamefont {L.}~\bibnamefont {Randall}}, \
  and\ \bibinfo {author} {\bibfnamefont {B.}~\bibnamefont {Shuve}},\ }\href
  {\doibase 10.1007/JHEP05(2011)097} {\bibfield  {journal} {\bibinfo  {journal}
  {JHEP}\ }\textbf {\bibinfo {volume} {05}},\ \bibinfo {pages} {097} (\bibinfo
  {year} {2011})},\ \Eprint {http://arxiv.org/abs/0909.4549}{arXiv:0909.4549
  [hep-ph]}\BibitemShut {NoStop}%
\bibitem [{\citenamefont {Asai}\ \emph {et~al.}(2007)\citenamefont {Asai},
  \citenamefont {Moroi}, \citenamefont {Nishihara},\ and\ \citenamefont
  {Yanagida}}]{Asai:2007sw}%
  \BibitemOpen
  \bibfield  {author} {\bibinfo {author} {\bibfnamefont {S.}~\bibnamefont
  {Asai}}, \bibinfo {author} {\bibfnamefont {T.}~\bibnamefont {Moroi}},
  \bibinfo {author} {\bibfnamefont {K.}~\bibnamefont {Nishihara}}, \ and\
  \bibinfo {author} {\bibfnamefont {T.~T.}\ \bibnamefont {Yanagida}},\ }\href
  {\doibase 10.1016/j.physletb.2007.06.080} {\bibfield  {journal} {\bibinfo
  {journal} {Phys. Lett.}\ }\textbf {\bibinfo {volume} {B653}},\ \bibinfo
  {pages} {81} (\bibinfo {year} {2007})},\ \Eprint
  {http://arxiv.org/abs/0705.3086}{arXiv:0705.3086 [hep-ph]}\BibitemShut
  {NoStop}%
\bibitem [{\citenamefont {Asai}\ \emph {et~al.}(2008)\citenamefont {Asai},
  \citenamefont {Moroi},\ and\ \citenamefont {Yanagida}}]{Asai:2008sk}%
  \BibitemOpen
  \bibfield  {author} {\bibinfo {author} {\bibfnamefont {S.}~\bibnamefont
  {Asai}}, \bibinfo {author} {\bibfnamefont {T.}~\bibnamefont {Moroi}}, \ and\
  \bibinfo {author} {\bibfnamefont {T.~T.}\ \bibnamefont {Yanagida}},\ }\href
  {\doibase 10.1016/j.physletb.2008.05.019} {\bibfield  {journal} {\bibinfo
  {journal} {Phys. Lett.}\ }\textbf {\bibinfo {volume} {B664}},\ \bibinfo
  {pages} {185} (\bibinfo {year} {2008})},\ \Eprint
  {http://arxiv.org/abs/0802.3725}{arXiv:0802.3725 [hep-ph]}\BibitemShut
  {NoStop}%
\bibitem [{\citenamefont {Asai}\ \emph {et~al.}(2009)\citenamefont {Asai},
  \citenamefont {Azuma}, \citenamefont {Jinnouchi}, \citenamefont {Moroi},
  \citenamefont {Shirai},\ and\ \citenamefont {Yanagida}}]{Asai:2008im}%
  \BibitemOpen
  \bibfield  {author} {\bibinfo {author} {\bibfnamefont {S.}~\bibnamefont
  {Asai}}, \bibinfo {author} {\bibfnamefont {Y.}~\bibnamefont {Azuma}},
  \bibinfo {author} {\bibfnamefont {O.}~\bibnamefont {Jinnouchi}}, \bibinfo
  {author} {\bibfnamefont {T.}~\bibnamefont {Moroi}}, \bibinfo {author}
  {\bibfnamefont {S.}~\bibnamefont {Shirai}}, \ and\ \bibinfo {author}
  {\bibfnamefont {T.~T.}\ \bibnamefont {Yanagida}},\ }\href {\doibase
  10.1016/j.physletb.2009.01.045} {\bibfield  {journal} {\bibinfo  {journal}
  {Phys. Lett.}\ }\textbf {\bibinfo {volume} {B672}},\ \bibinfo {pages} {339}
  (\bibinfo {year} {2009})},\ \Eprint
  {http://arxiv.org/abs/0807.4987}{arXiv:0807.4987 [hep-ph]}\BibitemShut
  {NoStop}%
\bibitem [{\citenamefont {{The ATLAS collaboration}}()}]{ATLAS:2017bna}%
  \BibitemOpen
  \bibfield  {author} {\bibinfo {author} {\bibnamefont {{The ATLAS
  collaboration}}},\ }\href@noop {} {\ }\Eprint
  {http://arxiv.org/abs/ATLAS-CONF-2017-017}{ATLAS-CONF-2017-017}\BibitemShut
  {NoStop}%
\bibitem [{\citenamefont {Ibe}\ \emph {et~al.}(2013)\citenamefont {Ibe},
  \citenamefont {Matsumoto},\ and\ \citenamefont {Sato}}]{Ibe:2012sx}%
  \BibitemOpen
  \bibfield  {author} {\bibinfo {author} {\bibfnamefont {M.}~\bibnamefont
  {Ibe}}, \bibinfo {author} {\bibfnamefont {S.}~\bibnamefont {Matsumoto}}, \
  and\ \bibinfo {author} {\bibfnamefont {R.}~\bibnamefont {Sato}},\ }\href
  {\doibase 10.1016/j.physletb.2013.03.015} {\bibfield  {journal} {\bibinfo
  {journal} {Phys. Lett.}\ }\textbf {\bibinfo {volume} {B721}},\ \bibinfo
  {pages} {252} (\bibinfo {year} {2013})},\ \Eprint
  {http://arxiv.org/abs/1212.5989}{arXiv:1212.5989 [hep-ph]}\BibitemShut
  {NoStop}%
\bibitem [{\citenamefont {McKay}\ \emph {et~al.}(2017)\citenamefont {McKay},
  \citenamefont {Scott},\ and\ \citenamefont {Athron}}]{McKay:2017rjs}%
  \BibitemOpen
  \bibfield  {author} {\bibinfo {author} {\bibfnamefont {J.}~\bibnamefont
  {McKay}}, \bibinfo {author} {\bibfnamefont {P.}~\bibnamefont {Scott}}, \ and\
  \bibinfo {author} {\bibfnamefont {P.}~\bibnamefont {Athron}},\ }\href@noop {}
  {\  (\bibinfo {year} {2017})},\ \Eprint
  {http://arxiv.org/abs/1710.01511}{arXiv:1710.01511 [hep-ph]}\BibitemShut
  {NoStop}%
\bibitem [{\citenamefont {Mahbubani}\ \emph {et~al.}(2017)\citenamefont
  {Mahbubani}, \citenamefont {Schwaller},\ and\ \citenamefont
  {Zurita}}]{Mahbubani:2017gjh}%
  \BibitemOpen
  \bibfield  {author} {\bibinfo {author} {\bibfnamefont {R.}~\bibnamefont
  {Mahbubani}}, \bibinfo {author} {\bibfnamefont {P.}~\bibnamefont
  {Schwaller}}, \ and\ \bibinfo {author} {\bibfnamefont {J.}~\bibnamefont
  {Zurita}},\ }\href {\doibase 10.1007/JHEP06(2017)119,
  10.1007/JHEP10(2017)061} {\bibfield  {journal} {\bibinfo  {journal} {JHEP}\
  }\textbf {\bibinfo {volume} {06}},\ \bibinfo {pages} {119} (\bibinfo {year}
  {2017})},\ \bibinfo {note} {[Erratum: JHEP10,061(2017)]},\ \Eprint
  {http://arxiv.org/abs/1703.05327}{arXiv:1703.05327 [hep-ph]}\BibitemShut
  {NoStop}%
\bibitem [{\citenamefont {Fukuda}\ \emph {et~al.}(2017)\citenamefont {Fukuda},
  \citenamefont {Nagata}, \citenamefont {Otono},\ and\ \citenamefont
  {Shirai}}]{Fukuda:2017jmk}%
  \BibitemOpen
  \bibfield  {author} {\bibinfo {author} {\bibfnamefont {H.}~\bibnamefont
  {Fukuda}}, \bibinfo {author} {\bibfnamefont {N.}~\bibnamefont {Nagata}},
  \bibinfo {author} {\bibfnamefont {H.}~\bibnamefont {Otono}}, \ and\ \bibinfo
  {author} {\bibfnamefont {S.}~\bibnamefont {Shirai}},\ }\href@noop {} {\
  (\bibinfo {year} {2017})},\ \Eprint
  {http://arxiv.org/abs/1703.09675}{arXiv:1703.09675 [hep-ph]}\BibitemShut
  {NoStop}%
\bibitem [{\citenamefont {Arkani-Hamed}\ \emph {et~al.}(2006)\citenamefont
  {Arkani-Hamed}, \citenamefont {Delgado},\ and\ \citenamefont
  {Giudice}}]{ArkaniHamed:2006mb}%
  \BibitemOpen
  \bibfield  {author} {\bibinfo {author} {\bibfnamefont {N.}~\bibnamefont
  {Arkani-Hamed}}, \bibinfo {author} {\bibfnamefont {A.}~\bibnamefont
  {Delgado}}, \ and\ \bibinfo {author} {\bibfnamefont {G.~F.}\ \bibnamefont
  {Giudice}},\ }\href {\doibase 10.1016/j.nuclphysb.2006.02.010} {\bibfield
  {journal} {\bibinfo  {journal} {Nucl. Phys.}\ }\textbf {\bibinfo {volume}
  {B741}},\ \bibinfo {pages} {108} (\bibinfo {year} {2006})},\ \Eprint
  {http://arxiv.org/abs/hep-ph/0601041}{arXiv:hep-ph/0601041
  [hep-ph]}\BibitemShut {NoStop}%
\bibitem [{\citenamefont {Harigaya}\ \emph {et~al.}(2014)\citenamefont
  {Harigaya}, \citenamefont {Kaneta},\ and\ \citenamefont
  {Matsumoto}}]{Harigaya:2014dwa}%
  \BibitemOpen
  \bibfield  {author} {\bibinfo {author} {\bibfnamefont {K.}~\bibnamefont
  {Harigaya}}, \bibinfo {author} {\bibfnamefont {K.}~\bibnamefont {Kaneta}}, \
  and\ \bibinfo {author} {\bibfnamefont {S.}~\bibnamefont {Matsumoto}},\ }\href
  {\doibase 10.1103/PhysRevD.89.115021} {\bibfield  {journal} {\bibinfo
  {journal} {Phys. Rev.}\ }\textbf {\bibinfo {volume} {D89}},\ \bibinfo {pages}
  {115021} (\bibinfo {year} {2014})},\ \Eprint
  {http://arxiv.org/abs/1403.0715}{arXiv:1403.0715 [hep-ph]}\BibitemShut
  {NoStop}%
\bibitem [{\citenamefont {Bharucha}\ \emph {et~al.}(2017)\citenamefont
  {Bharucha}, \citenamefont {Brummer},\ and\ \citenamefont
  {Ruffault}}]{Bharucha:2017ltz}%
  \BibitemOpen
  \bibfield  {author} {\bibinfo {author} {\bibfnamefont {A.}~\bibnamefont
  {Bharucha}}, \bibinfo {author} {\bibfnamefont {F.}~\bibnamefont {Brummer}}, \
  and\ \bibinfo {author} {\bibfnamefont {R.}~\bibnamefont {Ruffault}},\ }\href
  {\doibase 10.1007/JHEP09(2017)160} {\bibfield  {journal} {\bibinfo  {journal}
  {JHEP}\ }\textbf {\bibinfo {volume} {09}},\ \bibinfo {pages} {160} (\bibinfo
  {year} {2017})},\ \Eprint {http://arxiv.org/abs/1703.00370}{arXiv:1703.00370
  [hep-ph]}\BibitemShut {NoStop}%
\bibitem [{\citenamefont {Nagata}\ \emph {et~al.}(2015)\citenamefont {Nagata},
  \citenamefont {Otono},\ and\ \citenamefont {Shirai}}]{Nagata:2015pra}%
  \BibitemOpen
  \bibfield  {author} {\bibinfo {author} {\bibfnamefont {N.}~\bibnamefont
  {Nagata}}, \bibinfo {author} {\bibfnamefont {H.}~\bibnamefont {Otono}}, \
  and\ \bibinfo {author} {\bibfnamefont {S.}~\bibnamefont {Shirai}},\ }\href
  {\doibase 10.1007/JHEP10(2015)086} {\bibfield  {journal} {\bibinfo  {journal}
  {JHEP}\ }\textbf {\bibinfo {volume} {10}},\ \bibinfo {pages} {086} (\bibinfo
  {year} {2015})},\ \Eprint {http://arxiv.org/abs/1506.08206}{arXiv:1506.08206
  [hep-ph]}\BibitemShut {NoStop}%
\bibitem [{\citenamefont {Hisano}\ \emph
  {et~al.}(2015{\natexlab{b}})\citenamefont {Hisano}, \citenamefont
  {Kobayashi}, \citenamefont {Mori},\ and\ \citenamefont
  {Senaha}}]{Hisano:2014kua}%
  \BibitemOpen
  \bibfield  {author} {\bibinfo {author} {\bibfnamefont {J.}~\bibnamefont
  {Hisano}}, \bibinfo {author} {\bibfnamefont {D.}~\bibnamefont {Kobayashi}},
  \bibinfo {author} {\bibfnamefont {N.}~\bibnamefont {Mori}}, \ and\ \bibinfo
  {author} {\bibfnamefont {E.}~\bibnamefont {Senaha}},\ }\href {\doibase
  10.1016/j.physletb.2015.01.012} {\bibfield  {journal} {\bibinfo  {journal}
  {Phys. Lett.}\ }\textbf {\bibinfo {volume} {B742}},\ \bibinfo {pages} {80}
  (\bibinfo {year} {2015}{\natexlab{b}})},\ \Eprint
  {http://arxiv.org/abs/1410.3569}{arXiv:1410.3569 [hep-ph]}\BibitemShut
  {NoStop}%
\bibitem [{\citenamefont {Nagata}\ and\ \citenamefont
  {Shirai}(2015{\natexlab{a}})}]{Nagata:2014wma}%
  \BibitemOpen
  \bibfield  {author} {\bibinfo {author} {\bibfnamefont {N.}~\bibnamefont
  {Nagata}}\ and\ \bibinfo {author} {\bibfnamefont {S.}~\bibnamefont
  {Shirai}},\ }\href {\doibase 10.1007/JHEP01(2015)029} {\bibfield  {journal}
  {\bibinfo  {journal} {JHEP}\ }\textbf {\bibinfo {volume} {01}},\ \bibinfo
  {pages} {029} (\bibinfo {year} {2015}{\natexlab{a}})},\ \Eprint
  {http://arxiv.org/abs/1410.4549}{arXiv:1410.4549 [hep-ph]}\BibitemShut
  {NoStop}%
\bibitem [{\citenamefont {Nagata}\ and\ \citenamefont
  {Shirai}(2015{\natexlab{b}})}]{Nagata:2014aoa}%
  \BibitemOpen
  \bibfield  {author} {\bibinfo {author} {\bibfnamefont {N.}~\bibnamefont
  {Nagata}}\ and\ \bibinfo {author} {\bibfnamefont {S.}~\bibnamefont
  {Shirai}},\ }\href {\doibase 10.1103/PhysRevD.91.055035} {\bibfield
  {journal} {\bibinfo  {journal} {Phys. Rev.}\ }\textbf {\bibinfo {volume}
  {D91}},\ \bibinfo {pages} {055035} (\bibinfo {year} {2015}{\natexlab{b}})},\
  \Eprint {http://arxiv.org/abs/1411.0752}{arXiv:1411.0752
  [hep-ph]}\BibitemShut {NoStop}%
\bibitem [{\citenamefont {Alves}\ \emph {et~al.}(2015)\citenamefont {Alves},
  \citenamefont {Galloway}, \citenamefont {Ruderman},\ and\ \citenamefont
  {Walsh}}]{Alves:2014cda}%
  \BibitemOpen
  \bibfield  {author} {\bibinfo {author} {\bibfnamefont {D.~S.~M.}\
  \bibnamefont {Alves}}, \bibinfo {author} {\bibfnamefont {J.}~\bibnamefont
  {Galloway}}, \bibinfo {author} {\bibfnamefont {J.~T.}\ \bibnamefont
  {Ruderman}}, \ and\ \bibinfo {author} {\bibfnamefont {J.~R.}\ \bibnamefont
  {Walsh}},\ }\href {\doibase 10.1007/JHEP02(2015)007} {\bibfield  {journal}
  {\bibinfo  {journal} {JHEP}\ }\textbf {\bibinfo {volume} {02}},\ \bibinfo
  {pages} {007} (\bibinfo {year} {2015})},\ \Eprint
  {http://arxiv.org/abs/1410.6810}{arXiv:1410.6810 [hep-ph]}\BibitemShut
  {NoStop}%
\bibitem [{\citenamefont {Gross}\ \emph {et~al.}(2017)\citenamefont {Gross},
  \citenamefont {Lebedev},\ and\ \citenamefont {No}}]{Gross:2016ioi}%
  \BibitemOpen
  \bibfield  {author} {\bibinfo {author} {\bibfnamefont {C.}~\bibnamefont
  {Gross}}, \bibinfo {author} {\bibfnamefont {O.}~\bibnamefont {Lebedev}}, \
  and\ \bibinfo {author} {\bibfnamefont {J.~M.}\ \bibnamefont {No}},\ }\href
  {\doibase 10.1142/S0217732317500948} {\bibfield  {journal} {\bibinfo
  {journal} {Mod. Phys. Lett.}\ }\textbf {\bibinfo {volume} {A32}},\ \bibinfo
  {pages} {1750094} (\bibinfo {year} {2017})},\ \Eprint
  {http://arxiv.org/abs/1602.03877}{arXiv:1602.03877 [hep-ph]}\BibitemShut
  {NoStop}%
\bibitem [{\citenamefont {Farina}\ \emph {et~al.}(2017)\citenamefont {Farina},
  \citenamefont {Panico}, \citenamefont {Pappadopulo}, \citenamefont
  {Ruderman}, \citenamefont {Torre},\ and\ \citenamefont
  {Wulzer}}]{Farina:2016rws}%
  \BibitemOpen
  \bibfield  {author} {\bibinfo {author} {\bibfnamefont {M.}~\bibnamefont
  {Farina}}, \bibinfo {author} {\bibfnamefont {G.}~\bibnamefont {Panico}},
  \bibinfo {author} {\bibfnamefont {D.}~\bibnamefont {Pappadopulo}}, \bibinfo
  {author} {\bibfnamefont {J.~T.}\ \bibnamefont {Ruderman}}, \bibinfo {author}
  {\bibfnamefont {R.}~\bibnamefont {Torre}}, \ and\ \bibinfo {author}
  {\bibfnamefont {A.}~\bibnamefont {Wulzer}},\ }\href {\doibase
  10.1016/j.physletb.2017.06.043} {\bibfield  {journal} {\bibinfo  {journal}
  {Phys. Lett.}\ }\textbf {\bibinfo {volume} {B772}},\ \bibinfo {pages} {210}
  (\bibinfo {year} {2017})},\ \Eprint
  {http://arxiv.org/abs/1609.08157}{arXiv:1609.08157 [hep-ph]}\BibitemShut
  {NoStop}%
\bibitem [{\citenamefont {Aaboud}\ \emph {et~al.}(2017)\citenamefont {Aaboud}
  \emph {et~al.}}]{Aaboud:2017buh}%
  \BibitemOpen
  \bibfield  {author} {\bibinfo {author} {\bibfnamefont {M.}~\bibnamefont
  {Aaboud}} \emph {et~al.} (\bibinfo {collaboration} {ATLAS}),\ }\href@noop {}
  {\  (\bibinfo {year} {2017})},\ \Eprint
  {http://arxiv.org/abs/1707.02424}{arXiv:1707.02424 [hep-ex]}\BibitemShut
  {NoStop}%
\bibitem [{\citenamefont {Aaltonen}\ \emph {et~al.}(2009)\citenamefont
  {Aaltonen} \emph {et~al.}}]{Aaltonen:2008dn}%
  \BibitemOpen
  \bibfield  {author} {\bibinfo {author} {\bibfnamefont {T.}~\bibnamefont
  {Aaltonen}} \emph {et~al.} (\bibinfo {collaboration} {CDF}),\ }\href
  {\doibase 10.1103/PhysRevD.79.112002} {\bibfield  {journal} {\bibinfo
  {journal} {Phys. Rev.}\ }\textbf {\bibinfo {volume} {D79}},\ \bibinfo {pages}
  {112002} (\bibinfo {year} {2009})},\ \Eprint
  {http://arxiv.org/abs/0812.4036}{arXiv:0812.4036 [hep-ex]}\BibitemShut
  {NoStop}%
\bibitem [{\citenamefont {Harigaya}\ \emph {et~al.}(2015)\citenamefont
  {Harigaya}, \citenamefont {Ichikawa}, \citenamefont {Kundu}, \citenamefont
  {Matsumoto},\ and\ \citenamefont {Shirai}}]{Harigaya:2015yaa}%
  \BibitemOpen
  \bibfield  {author} {\bibinfo {author} {\bibfnamefont {K.}~\bibnamefont
  {Harigaya}}, \bibinfo {author} {\bibfnamefont {K.}~\bibnamefont {Ichikawa}},
  \bibinfo {author} {\bibfnamefont {A.}~\bibnamefont {Kundu}}, \bibinfo
  {author} {\bibfnamefont {S.}~\bibnamefont {Matsumoto}}, \ and\ \bibinfo
  {author} {\bibfnamefont {S.}~\bibnamefont {Shirai}},\ }\href {\doibase
  10.1007/JHEP09(2015)105} {\bibfield  {journal} {\bibinfo  {journal} {JHEP}\
  }\textbf {\bibinfo {volume} {09}},\ \bibinfo {pages} {105} (\bibinfo {year}
  {2015})},\ \Eprint {http://arxiv.org/abs/1504.03402}{arXiv:1504.03402
  [hep-ph]}\BibitemShut {NoStop}%
\end{thebibliography}%

\end{document}